\documentclass[aps,prl,notitlepage,twocolumn,superscriptaddress,10pt,floatfix]{revtex4-2}

\makeatletter
\def\@hangfrom@section#1#2#3{\normalsize\@hangfrom{#1#2}#3}
\def\@hangfroms@section#1#2{\normalsize#1#2}
\makeatother

\usepackage{amsmath,amsfonts,amssymb,bm}
\usepackage{graphicx}
\usepackage{enumerate}

\usepackage{theorem}

\usepackage{tikz}

\definecolor{darkblue1}{rgb}{0.18,0.19,0.57}
\usepackage[colorlinks=true, allcolors=darkblue1]{hyperref}

\newcommand{\nc}{\newcommand}

\nc{\ket}[1]{|#1\rangle}
\nc{\bra}[1]{\langle#1|}
\nc{\ketbra}[2]{|#1\rangle\!\langle#2|}
\nc{\braket}[2]{\langle#1|#2\rangle}
\nc{\braoprket}[3]{\langle#1|#2|#3\rangle}
\nc{\opr}[1]{\operatorname{#1}}
\nc{\avg}[1]{\langle#1\rangle}
\nc{\ketbrasame}[1]{|#1\rangle\!\langle#1|}
\nc{\E}{\mathbb{E}}
\nc{\var}{\operatorname{Var}}

\nc{\hk}[1]{\textcolor{violet}{\textbf{[hk: #1]}}}
\nc{\ch}[1]{\textcolor{purple}{\textbf{[ch: #1]}}}
\nc{\ls}[1]{\textcolor{blue}{\textbf{[ls: #1]}}}
\usepackage[normalem]{ulem}
\nc{\sxz}[1]{\textcolor{violet}{sxz:#1}}

\begin{document}
\title{Persistent Nonlocal Magic Revivals from Entanglement-Spectrum Broadening}
\author{Lv Zhang}
\email{zhanglv@iphy.ac.cn}
\affiliation{Beijing National Laboratory for Condensed Matter Physics,
Institute of Physics, Chinese Academy of Sciences, Beijing 100190, China}
\affiliation{School of Physical Sciences, University of Chinese Academy of Sciences, Beijing 100049, China}

\author{Tian-Ren Jin}
\email{jintianren@sztu.edu.cn}
\affiliation{Shenzhen Key Laboratory of Ultraintense Laser and Advanced Material Technology, Center for Intense Laser Application Technology, and College of Engineering Physics, Shenzhen Technology University, Shenzhen 518118, China}
\affiliation{College of Applied Sciences, Shenzhen University, Shenzhen, 518060, China}

\author{Heng Fan}
\email{hfan@iphy.ac.cn}
\affiliation{Beijing National Laboratory for Condensed Matter Physics,
Institute of Physics, Chinese Academy of Sciences, Beijing 100190, China}
\affiliation{School of Physical Sciences, University of Chinese Academy of Sciences, Beijing 100049, China}
\affiliation{Beijing Key Laboratory of Advanced Quantum Technology,
Beijing Academy of Quantum Information Sciences, Beijing 100193, China}
\affiliation{Hefei National Laboratory, Hefei 230088, China}
\affiliation{Songshan Lake Materials Laboratory, Dongguan, Guangdong 523808, China}

\date{\today}

\begin{abstract}
Min-relative nonlocal magic (NLM) quantifies nonstabilizerness that  cannot be removed by local unitaries. Using qubit stabilizer states as the reference, we find asymptotically persistent min-relative NLM revivals under two-mode squeezing despite growing entanglement and a decreasing return probability. By considering Schmidt spectra that broaden while retaining their shape, we show that persistent NLM revivals can arise, with the spectral shape setting their asymptotic amplitude and the broadening law governing their temporal recurrence. We verify the predicted amplitude and asymptotic doubling of successive revival times in the early broadening regime of number-conserving two-mode mixing at large particle number.
\end{abstract}

\maketitle

\textit{Introduction.---}
Entanglement growth is a major challenge for tensor-network simulations of quantum many-body dynamics~\cite{Schuch_2008,PhysRevLett.91.147902,PhysRevLett.93.040502,PhysRevLett.100.030504}. Nevertheless, Clifford circuits acting on stabilizer states can generate extensive entanglement while remaining efficiently simulable on classical computers~\cite{PhysRevA.74.062314,PhysRevX.7.031016,PhysRevA.70.052328}. This distinction motivates a complementary resource perspective based on magic, or nonstabilizerness, which characterizes departures from the stabilizer framework and supplies an essential ingredient for universal quantum computation~\cite{PhysRevLett.128.050402,Veitch_2014,PhysRevLett.118.090501,PhysRevA.71.022316,Reichardt2005,PRXQuantum.3.020333,PhysRevLett.116.250501,PhysRevA.86.052329}. Understanding how magic is generated and evolves is therefore important for characterizing quantum resources in nonequilibrium systems~\cite{goto2021chaosmagic,Turkeshi2025,PhysRevB.106.125130,njgn-fksh,79vj-nx6r,c98q-fhbn,xfp5-hhs4,h7vb-gc5d}. Local non-Clifford rotations can change magic without changing bipartite entanglement, raising the question of how much nonstabilizerness is intrinsic to correlations between subsystems~\cite{PhysRevB.107.035148,ahmad2026experimentaldemonstrationnonlocalmagic}. Nonlocal magic (NLM) addresses this question by minimizing magic over arbitrary local unitaries on the two sides of a bipartition~\cite{Cao2024,z3vr-w5c5,Qian_2025,torre2026nonlocalmagicentanglementspectrum,liu2026nonlocalnonstabilizernessslightlyentangled,liu2026entanglementantiflatnessnonlocalnonstabilizerness,9x56-2b45,jh55-rmc6,huang2026intrinsicspectralstructurebipartite,torre2026nonlocalmagicentanglementspectrum,sierant2026spectralgeometrynonlocalstabilizer}. Its dynamics probes how nonstabilizer structure that cannot be removed by these local unitaries generates and evolves as quantum correlations are established and reorganized.

Min-relative NLM is defined as the negative logarithm of the largest stabilizer fidelity attainable under local unitaries~\cite{PRXQuantum.3.020333,sierant2026exactquantificationnonlocalmagic,viscardi2026nonlocalmagicclosedformsolution,shaposhnik2026infinitemagicemergencecontinuum}. Recent results show that, for bipartite pure qubit states, this optimization reduces to maximizing the fidelity between the Schmidt spectrum and flat reference spectra of dyadic rank $2^k$~\cite{sierant2026exactquantificationnonlocalmagic,viscardi2026nonlocalmagicclosedformsolution,ebner2026globalnonlocalmagicquantum}. Min-relative NLM therefore quantifies the departure of the evolving Schmidt spectrum from its best match among these dyadic-flat reference spectra. This characterization makes NLM directly accessible from the entanglement spectrum and provides a framework for investigating how the redistribution of Schmidt weights governs NLM dynamics~\cite{zhang2026entanglementgrowthtransportschmidt,minar2026magicbarrieruniversalfeature,li2026nonlocalmagicmanybodylocalization,karjula2026universalentanglementembezzlementdivergent,aditya2026nonlocalmagicspreadingmanybody,karjula2026magicalmemorycostentanglement}. 

Recent studies have revealed distinct dynamical behaviors of min-relative NLM in many-body systems. In random circuits and chaotic Hamiltonian quenches, NLM exhibits an initial buildup followed by a transient maximum and relaxation toward a stationary value~\cite{karjula2026universalentanglementembezzlementdivergent,aditya2026nonlocalmagicspreadingmanybody}. Dimer-state quenches in free-fermion chains instead show growth followed by saturation, with the stationary NLM scaling logarithmically with subsystem size~\cite{aditya2026nonlocalmagicspreadingmanybody}. N\'eel-state quenches of the random-field XXZ chain further reveal disorder-dependent behavior, with overshoot and relaxation at weak disorder and slow, approximately logarithmic NLM growth at strong disorder over the accessible time window~\cite{li2026nonlocalmagicmanybodylocalization,zhang2026entanglementgrowthtransportschmidt}. For disorder-averaged N\'eel-state quenches with periodic boundary conditions, Zhang et al.~\cite{zhang2026entanglementgrowthtransportschmidt} identified a temporal hierarchy: the first NLM peak follows the maximum rate of entanglement production and precedes the time at which dominant-Schmidt-scale transport becomes typical. These results connect min-relative NLM dynamics to the redistribution of Schmidt weights and reveal spectral reorganization beyond that captured by entanglement growth alone.

These results motivate asking whether min-relative NLM can exhibit asymptotically persistent revivals even as entanglement grows and the initial-state return probability decreases. The fidelity optimization compares a continuously evolving Schmidt spectrum with flat reference spectra whose allowed ranks form the discrete sequence $2^k$. This structure motivates examining spectra that broaden while approaching a fixed rescaled shape, and asking whether their matching to successive reference sectors can repeatedly reproduce the same pattern. The central questions are what allows these revivals to retain a nonzero asymptotic amplitude and how the spectral shape and broadening law determine their temporal recurrence.

In this work, we find asymptotically persistent min-relative NLM revivals under two-mode squeezing despite growing entanglement and a decreasing initial-state return probability. Throughout, we quantify min-relative NLM in the bosonic models relative to the qubit stabilizer structure through local qubit encodings. For infinite-dimensional squeezing, we establish a controlled limit of increasing qubit register size. Finite-dimensional squeezing reproduces the ideal revival pattern over a time window that expands with local dimension,  encompassing progressively more revival cycles. We then establish a general spectral mechanism: for Schmidt spectra that broaden toward a fixed rescaled shape, successive dyadic fidelity branches approach translated copies of a common profile on the logarithmic-width axis. The spectral shape fixes the limiting optimized profile and its amplitude, while the broadening law determines the temporal recurrence whenever the profile is nonconstant. Finally, in number-conserving two-mode mixing at large particle number, we verify the predicted amplitude and asymptotic doubling of successive revival times within the controlled early broadening regime.

\textit{Two-mode squeezing.---}
We consider two bosonic modes $A$ and $B$ undergoing two-mode
squeezing generated~\cite{PhysRevA.31.3093} by the Hamiltonian
\begin{equation}
H=i\chi(a^\dagger b^\dagger-ab),
\label{eq:squeezing-hamiltonian}
\end{equation}
where $a$ and $b$ are the annihilation operators for modes $A$
and $B$, respectively, and $a^\dagger$ and $b^\dagger$ are the
corresponding creation operators.
We take $\chi>0$ as the squeezing rate.
Starting from the joint vacuum state
$|0,0\rangle=|0\rangle_A|0\rangle_B$, the evolution produces
the two-mode squeezed state
\begin{equation}
|\Psi(r)\rangle=e^{-iHt}|0,0\rangle
=\frac{1}{\cosh r}\sum_{m=0}^{\infty}
(\tanh r)^m|m,m\rangle,
\label{eq:squeezed-state}
\end{equation}
where $r=\chi t$ is the dimensionless squeezing parameter and
$|m,m\rangle=|m\rangle_A|m\rangle_B$, with $|m\rangle_{A,B}$
denoting local Fock states.
This is already a Schmidt decomposition, whose probabilities follow the geometric distribution
\begin{equation}
\lambda_m=(1-q)q^m,\qquad q=\tanh^2 r.
\label{eq:geometric-spectrum}
\end{equation}
This geometric Schmidt spectrum broadens continuously as
$q\to1^-$, with mean occupation
$\bar n=q/(1-q)=\sinh^2 r$ in each mode.

\begin{figure}[t]
\centering
 \includegraphics[width=1.0\linewidth]{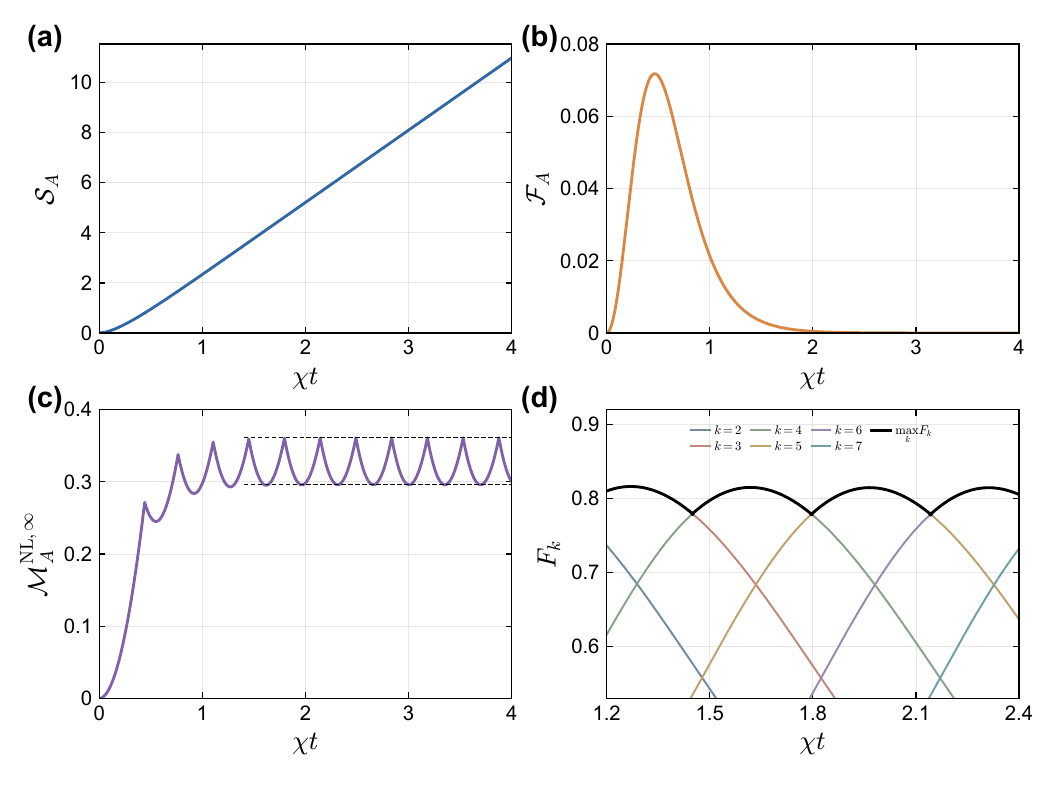}
\caption{Spectral diagnostics of two-mode squeezing as functions of
$r=\chi t$, with Schmidt probabilities $\lambda_m=(1-q)q^m$ and
$q=\tanh^2 r$. (a) The entanglement entropy
$\mathcal{S}_A$ grows asymptotically
linearly. (b) The spectral anti-flatness
$\mathcal{F}_A=$ rises initially and then decays to zero.
(c) The min-relative NLM
$\mathcal M_A^{\mathrm{NL},\infty}$
exhibits persistent oscillations, its asymptotic period is
$T=\ln 2/(2\chi)$ and the peak-to-trough
revival amplitude approaches $0.065$, between two dashed lines. (d) The dyadic fidelity branches
$F_k$
and their upper envelope (black line) show how successive changes in the
optimal Bell-pair sector produce the oscillations in (c).}
\label{fig1}
\end{figure}

\begin{figure}[t]
\centering
 \includegraphics[width=1.0\linewidth]{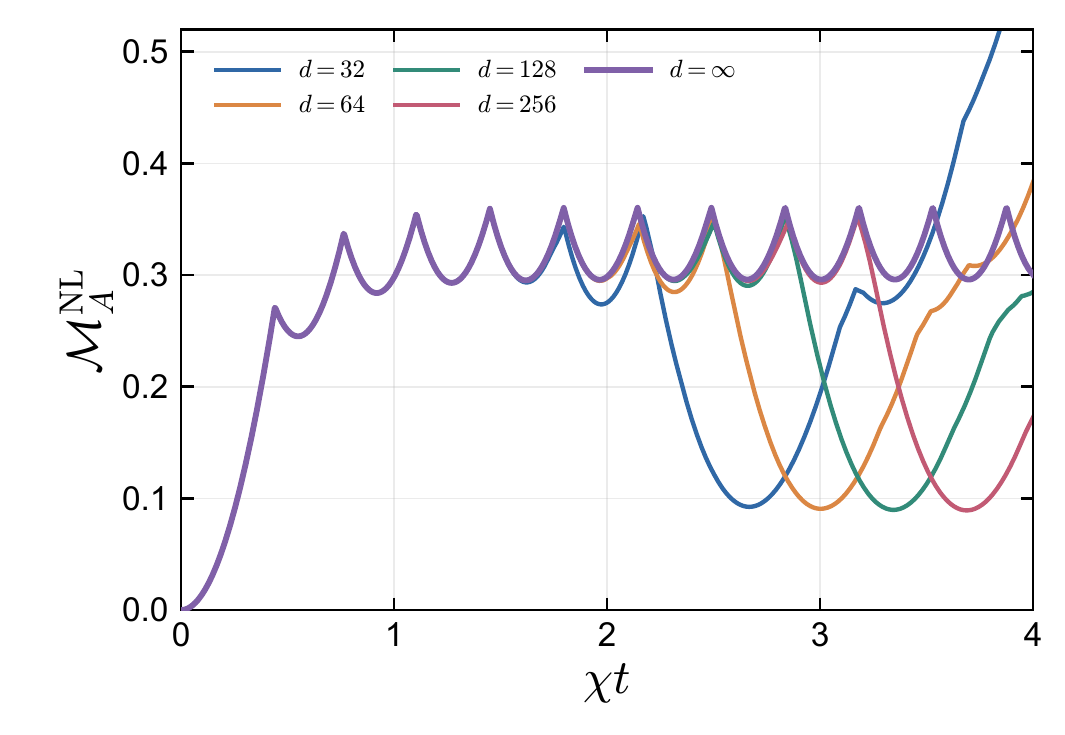}
\caption{Finite-dimensional effects on two-mode squeezing.
The locally optimized min-relative NLM
$\mathcal M_A^{\mathrm{NL},\infty}$ as a function of $r=\chi t$,
starting from the joint vacuum.
For $d=32,64,128,$ and $256$, each mode retains the Fock levels
$0,\ldots,d-1$, and the state evolves under $H_d=P_dHP_d$,
where $H=i\chi(a^\dagger b^\dagger-ab)$ and $P_d$ projects
onto the retained two-mode space.
The purple curve shows the infinite-dimensional result.
The finite-dimensional curves initially reproduce its revival
pattern; increasing $d$ delays cutoff-induced deviations
and extends the interval of agreement.}
\label{fig2}
\end{figure}

We characterize the reduced state
$\rho_A=\operatorname{Tr}_B|\Psi(r)\rangle\langle\Psi(r)|$
through its entanglement entropy $\mathcal S_A$ and
spectral anti-flatness $\mathcal F_A$, a complementary
spectral diagnostic linked to nonstabilizerness~\cite{PhysRevA.109.L040401,z3vr-w5c5,wtq7-qy11,zhang2026revealingentanglementgrowthmechanismsmagic,liu2026entanglementantiflatnessnonlocalnonstabilizerness,jasser2026journeyflatlanddoesantiflatness}.
For this spectrum, both admit closed-form expressions:
\begin{align}
\mathcal S_A
&=-\operatorname{Tr}(\rho_A\log_2\rho_A)
\nonumber\\
&=-\log_2(1-q)-\frac{q}{1-q}\log_2 q,
\label{eq:entropy}\\
\mathcal F_A
&=\operatorname{Tr}(\rho_A^3)
-[\operatorname{Tr}(\rho_A^2)]^2
\nonumber\\
&=\frac{q(1-q)^2}{(1+q)^2(1+q+q^2)}.
\label{eq:antiflatness}
\end{align}
As shown in Fig.~\ref{fig1}(a), (b), the entanglement entropy grows
asymptotically linearly,
$\mathcal S_A=(2r+1)\log_2 e-2+o(1)$ as $r\to\infty$,
whereas the anti-flatness initially rises and subsequently
decays to zero as $\mathcal F_A\sim(4/3)e^{-4r}$. 

For the bosonic state, we retain the qubit stabilizer structure
as the resource reference and construct the infinite-dimensional
quantity through a controlled finite-register limit.
We first retain the Fock levels $m=0,\ldots,2^n-1$ in each
mode and normalize the Schmidt probabilities:
\begin{equation}
\lambda_m^{(n)}
=\frac{(1-q)q^m}{1-q^{2^n}},
\qquad 0\leq m<2^n.
\label{eq:truncated-spectrum}
\end{equation}
The resulting state is
$|\Psi_n\rangle=\sum_{m=0}^{2^n-1}
\sqrt{\lambda_m^{(n)}}|m,m\rangle$.
A local binary encoding,
$|m\rangle\mapsto|\operatorname{bin}_n(m)\rangle$,
identifies each truncated Fock space with an $n$-qubit
register while preserving the Schmidt spectrum.
Using this identification, we define the min-relative NLM
of $|\Psi_n\rangle$ across the two registers by
\begin{equation}
\mathcal M_A^{\mathrm{NL},(n)}
=-\log_2\!\left[
\max_{U_A,U_B}
F_{\mathrm{stab}}\bigl((U_A\otimes U_B)|\Psi_n\rangle\bigr)
\right],
\label{eq:finite-nlm-definition}
\end{equation}
where $F_{\mathrm{stab}}(|\phi\rangle)
=\max_{|s\rangle\in\mathrm{STAB}_{2n}}
|\langle s|\phi\rangle|^2$ is the fidelity with the closest
pure $2n$-qubit stabilizer state~\cite{Qian_2025,sierant2026exactquantificationnonlocalmagic}.
The optimization runs over arbitrary unitaries on each local register.

Applying the finite-dimensional spectral formula~\cite{sierant2026exactquantificationnonlocalmagic,viscardi2026nonlocalmagicclosedformsolution} to the truncated geometric spectrum gives
\begin{equation}
\mathcal M_A^{\mathrm{NL},(n)}
=-\log_2\!\left[
\frac{\max_{0\leq k\leq n}F_k(r)}{1-q^{2^n}}
\right],
\label{eq:finite-nlm-spectrum}
\end{equation}
where
\begin{equation}
F_k(r)=\frac{1}{2^k}
\left(
\sum_{m=0}^{2^k-1}\sqrt{(1-q)q^m}
\right)^2.
\label{eq:fidelity-branches}
\end{equation}
Here, $F_k$ is evaluated using the original, untruncated
probabilities.
Each branch represents the best fidelity with a state locally
equivalent to $k$-Bell pairs and product spectators.
Full optimization over local unitaries makes the result
independent of the local basis used for the encoding.

We define the spectral extension of qubit min-relative NLM
to the bosonic Schmidt spectrum through the finite-register
limit
\begin{align}
\mathcal M_A^{\mathrm{NL},\infty}(r)
&:=\lim_{n\to\infty}\mathcal M_A^{\mathrm{NL},(n)}(r)
\nonumber\\
&=-\log_2\max_{k\geq0}F_k(r),
\label{eq:infinite-nlm-definition}
\end{align}
where the limit is taken at fixed finite $r$.
The qubit stabilizer structure remains the resource reference,
and the resulting quantity depends only on the Schmidt
spectrum, independently of the local encoding.
At such fixed finite $r$, $q<1$, the normalized
truncated states converge to the squeezed state, with
$|\langle\Psi(r)|\Psi_n(r)\rangle|^2
=1-q^{2^n}\to1$ as $n\to\infty$.
To control the maximization over $k$, geometric summation
gives $F_k(r)\le e^{2r}2^{-k}\to0$ as $k\to\infty$,
while $F_0(r)=1-q>0$.
The maximum is therefore attained at a finite index $k_\ast$.
Once $n\ge k_\ast$, the truncated optimization includes
this optimal branch.
Since $1-q^{2^n}\to1$, the optimized fidelity converges
to $F_\ast(r)=\max_{k\ge0}F_k(r)$.
Moreover, $1-q\le F_\ast(r)\le1$, so its negative logarithm
has a finite limit:
\begin{equation}
0\le\mathcal M_A^{\mathrm{NL},\infty}(r)
=-\log_2F_\ast(r)
\le-\log_2(1-q)<\infty.
\end{equation}
Furthermore, zero-padding a finite spectrum into larger qubit
registers leaves its optimized min-relative  NLM unchanged.
These properties establish a consistent extension of the
locally optimized qubit-stabilizer measure to the bosonic
Schmidt spectrum.
Further details are provided in
Sec.~\ref{app:infinite-limit} of the Supplemental Material.

Fig.~\ref{fig1}(c) shows this
persistent min-relative NLM revivals.
Their origin is resolved by the fidelity branches in
Fig.~\ref{fig1}(d), whose adjacent ratios satisfy
\begin{equation}
\frac{F_{k+1}(r)}{F_k(r)}
=\frac{\left[1+(\tanh r)^{2^k}\right]^2}{2}.
\label{eq:branch-ratio}
\end{equation}
The optimal sector therefore changes from $k$ to $k+1$ at
\begin{equation}
r_k=\operatorname{arctanh}
\!\left[(\sqrt2-1)^{2^{-k}}\right],
\label{eq:sector-switch}
\end{equation}
producing the cusps in the min-relative NLM curve.

The asymptotic oscillations follow directly from this dyadic
structure. For $z=2^k e^{-2r}$ held fixed, the fidelity branches
approach
\begin{equation}
F_k(r)=h(z)+O(e^{-4r}),
\qquad
h(z)=\frac{(1-e^{-2z})^2}{z}.
\label{eq:asymptotic-branch}
\end{equation}
Define the asymptotic profile
\begin{equation}
Q(s)=-\log_2\!\left[
\max_{j\in\mathbb Z}h(2^{j-s})
\right],
\label{eq:periodic-profile}
\end{equation}
then the optimized min-relative NLM
can be written as
\begin{equation}
\mathcal M_A^{\mathrm{NL},\infty}(r)
=Q(2r\log_2e)+O(e^{-4r}).
\label{eq:asymptotic-nlm}
\end{equation}
Reindexing $j$ gives $Q(s+1)=Q(s)$, establishing
the periodicity of the limiting profile:
\begin{equation}
T=\frac{1}{2\chi\log_2 e}
=\frac{\ln2}{2\chi}.
\label{eq:nlm-period}
\end{equation}
The extrema of $Q(s)$ follow from the maximum of $h(z)$
and the crossing of adjacent dyadic branches,
$h(z)=h(2z)$.
These give $Q_{\min}\simeq0.296$ and
$Q_{\max}\simeq0.361$, yielding the nonzero
revival amplitude
\begin{equation}
\Delta\mathcal M=Q_{\max}-Q_{\min}\simeq0.065.
\label{eq:nlm-amplitude}
\end{equation}
The two dashed lines in Fig.~\ref{fig1}(c) show these
predicted asymptotic bounds, which agree with the
late-time extrema of the min-relative NLM.
These min-relative NLM revivals do not correspond to revivals of the
quantum state: the vacuum return probability,
$|\langle0,0|\Psi(r)\rangle|^2=\operatorname{sech}^2r$,
decreases monotonically, while the entanglement entropy
continues to grow.

Finally, Fig.~\ref{fig2} examines the dynamical
effects of a finite local dimension $d=2^n$.
We evolve the initial vacuum state directly under
$H_d=i\chi(a_d^\dagger b_d^\dagger-a_d b_d)$,
where $a_d$ and $b_d$ are the annihilation matrices restricted
to the local Fock levels $0,\ldots,d-1$, and evaluate min-relative NLM
from the decreasingly ordered Schmidt probabilities.
This evolution differs from the normalized projection
$|\Psi_n\rangle$ used to define the infinite-dimensional limit.
Fig.~\ref{fig2} shows that finite systems reproduce the
infinite-dimensional revival pattern over a time window
that grows with $d$. At later times, the Fock cutoff
modifies the dynamics and causes deviations.

\textit{Spectral broadening and NLM revivals.---}
The squeezing mechanism raises two questions: when does
a evolving spectrum repeat the same pattern, and
when does that pattern have a nonzero amplitude?
We address the first by considering spectra whose shape
is preserved after rescaling their rank axis by a growing
width $L(t)$. The probabilities then spread over more
levels while retaining the same relative distribution.
Let $f(u)$ be nonnegative, nonincreasing, and normalized
on $[0,\infty)$. We require the ordered Schmidt
probabilities $\{\lambda_j^{\downarrow}(t)\}$ to approach its cell-integrated spectrum:
\begin{align}
p_j^{(L)}
=\int_{j/L}^{(j+1)/L}f(u)\,du,
\label{eq:scale-spectrum}\\
\sum_{j\ge0}
\left[\sqrt{\lambda_j^\downarrow(t)}
-\sqrt{p_j^{(L(t))}}\right]^2\longrightarrow0,
\label{eq:scale-convergence}
\end{align}
with $L(t)\to\infty$.
The cell integral preserves discrete normalization exactly.
For large $L(t)$, the spectrum has the approximate scaling form
\begin{equation}
    \lambda_j^\downarrow(t)\approx
\frac{1}{L(t)}
f\!\left(\frac{j}{L(t)}\right).
\end{equation}

For a candidate stabilizer rank $2^k$, the ratio
$y=2^k/L$ measures how far its support extends across the
broadened spectrum. The limiting fidelity is
\begin{equation}
H_f(y)=\frac{1}{y}
\left(\int_0^y\sqrt{f(u)}\,du\right)^2.
\label{eq:general-profile}
\end{equation}
It is the squared overlap between the spectral amplitude
$\sqrt f$ and a normalized flat amplitude on $[0,y]$.
Sampling this function at the allowed dyadic ranks gives
$Q_f(s)=-\log_2\max_{\ell\in\mathbb Z}H_f(2^{\ell-s})$,
and the optimized NLM satisfies
\begin{align}
\mathcal M_A^{\mathrm{NL},\infty}(t)
&=Q_f(\log_2L(t))+o(1),
\nonumber\\
Q_f(s+1)&=Q_f(s).
\label{eq:general-scale-revivals}
\end{align}
Doubling the width $L\rightarrow 2L$ shifts the competing sectors by one and
leaves their optimized fidelity unchanged asymptotically.
The entire limiting profile is therefore predicted from
$f$, without fitting the min-relative NLM curve.

If $H_f$ has a unique global maximum, some dyadic phases
hit this optimum while others miss it, ensuring a nonzero
revival amplitude. A maximum plateau spanning $[a,2a]$,
however, is sampled at every phase and gives a constant
profile. Such a plateau can occur even for a strictly
decreasing spectral shape: broadening and increasing
entanglement alone do not guarantee persistent revivals.
For strictly unimodal $H_f$, this distinction becomes
quantitative. Let $y_m$ be its maximum and let
$y_c\in(y_m/2,y_m)$ satisfy $H_f(y_c)=H_f(2y_c)$.
The best match at $y_m$ gives the min-relative NLM minimum, while the
crossing of adjacent sectors at $y_c$ gives its maximum.
Their difference is
\begin{equation}
\Delta\mathcal M_f
=\log_2\frac{H_f(y_m)}{H_f(y_c)}>0.
\label{eq:general-revival-amplitude}
\end{equation}
The general nonconstant-profile criterion, counterexamples,
and the single-peak result are given in
Sec.~\ref{app:revival-criterion} of the Supplemental Material.

For a nonconstant profile, continuous growth through
successive scale doublings produces persistent revivals.
Their time dependence follows from the broadening law.
When $L(t)\sim L_0e^{\gamma t}$, each doubling takes the
same asymptotic interval $T=\ln2/\gamma$.
When $L(t)\sim L_0(t/t_0)^\nu$, the same phase instead
recurs at times separated by a factor $2^{1/\nu}$,
giving periodicity in logarithmic time.
These are fundamental periods when $H_f$ is strictly
unimodal. For squeezing, the geometric spectrum is the
exact discretization of $f(u)=e^{-u}$ with
$L=-1/\ln q\sim e^{2\chi t}/4$.
The general construction thus recovers both the period and
the finite revival amplitude observed in Fig.~\ref{fig1}.

\begin{figure}[t]
\centering
 \includegraphics[width=1.0\linewidth]{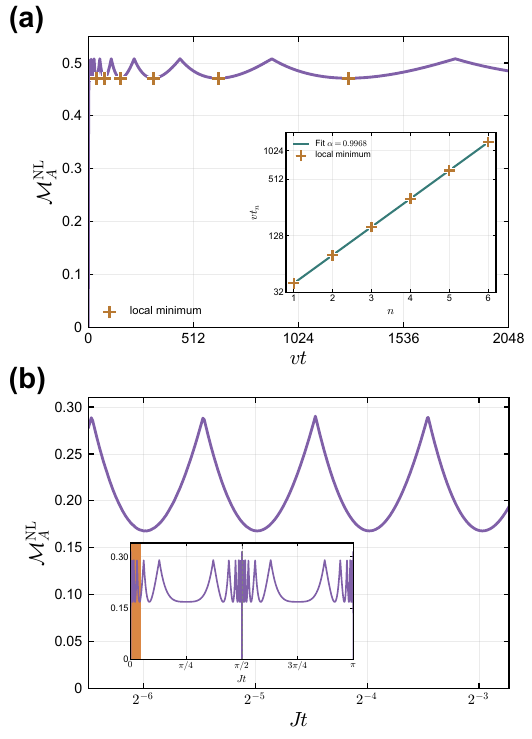}
\caption{Logarithmic-time min-relative NLM revivals.
(a) The min-relative NLM
$\mathcal M_A^{\mathrm{NL},\infty}$ as a function of $vt$
for the Schmidt spectrum
$\lambda_j(L)=(j+L)^{-3}/\zeta(3,L)$,
with $L=vt$ and $\zeta$ is the Hurwitz zeta function.
Crosses mark local minima.
The inset shows their times $vt_n$ as a function of the
minimum index $n$, on a logarithmic vertical scale.
The fitted slope of $\log_2(vt_n)$ versus $n$ is $0.9968$,
consistent with $t_{n+1}/t_n\to2$.
(b) The min-relative NLM as a function of $Jt$, on a logarithmic
horizontal scale, for number-conserving two-mode mixing
generated by $H=J(a^\dagger b+ab^\dagger)$,
starting from $|N,0\rangle$ with $N=2^{20}-1$.
The early-time oscillations are approximately periodic
in $\log_2(Jt)$.
The inset shows the NLM as a function of $Jt$ over the
full interval $0\leq Jt\leq\pi$; the orange shading
identifies the window displayed in the main panel.}
\label{fig3}
\end{figure}

\textit{Logarithmic-time revivals.---}
Exponential broadening makes the squeezing revivals equally
spaced at late times. Fig.~\ref{fig3} explores linear
broadening, for which the same recurrence in spectral width
produces increasing temporal separations.
We first prescribe the shifted inverse-cubic spectrum
\begin{equation}
\lambda_j(L)=\frac{(j+L)^{-3}}{\zeta(3,L)},
\qquad L=vt>0,
\label{eq:inverse-cubic-spectrum}
\end{equation}
where $\zeta$ is the Hurwitz zeta function.
Its limiting shape $f_{\mathrm P}(u)=2/(1+u)^3$ yields
a nonconstant profile with a smaller amplitude $\Delta \mathcal{M}\simeq 0.037 $ than the
exponential shape. Linear growth of $L$ predicts that
corresponding revival times asymptotically double $t_{n+1}/t_n=2$.
Fig.~\ref{fig3}(a) displays the growing separation of
successive minima; the near-unit slope in the inset
confirms their predicted spacing on the logarithmic-time
axis. The spectral limit and amplitude are derived in
Sec.~\ref{app:inverse-cubic-spectrum} of the Supplemental Material.

To realize this temporal law dynamically, we consider
number-conserving mixing between two bosonic modes $A$
and $B$, generated by
\begin{equation}
    H=J(a^\dagger b+ab^\dagger).
\end{equation}
Here, $a$ and $b$ are the annihilation operators of the
respective modes, $a^\dagger$ and $b^\dagger$ are their
creation operators, and $J>0$ is the coupling strength.
The Hamiltonian transfers bosons between the modes while
conserving their total occupation.

We initialize the system in
$|N,0\rangle=|N\rangle_A|0\rangle_B$,
with all $N$ bosons in mode $A$.
Conservation of total particle number confines the
evolution to the states $|N-m,m\rangle$, with
$0\le m\le N$.
Each mode therefore explores only the Fock levels
$0,\ldots,N$.
We choose $N+1=2^n$ so that each accessible local space
can be encoded in an $n$-qubit register.
The Schmidt probabilities follow the binomial law~\cite{PhysRevA.65.032323}:
\begin{equation}
    \lambda_m^{(N)}(t)=\binom{N}{m}p(t)^m[1-p(t)]^{N-m},
\end{equation}
with $p(t)=\sin^2(Jt)$.
In the broad-spectrum limit $Np(t)[1-p(t)]\to\infty$, the Schmidt
probabilities, sorted in decreasing order, approach the half-Gaussian
profile $f_{\mathrm G}(u)=2e^{-u^2}/\sqrt{\pi}$ on the rescaled rank
axis $u=j/L(t)$, where $L(t)=\sqrt{8Np(t)[1-p(t)]}$.
In the early-time window $N^{-1/2}\ll Jt\ll1$, this width grows
linearly, $L(t)\simeq\sqrt{8N}\,Jt$.
The model consequently has logarithmic-time revivals,
with a larger limiting amplitude $\Delta \mathcal{M}\simeq 0.122$ set by the half-Gaussian
shape. Fig.~\ref{fig3}(b) resolves them during the
initial broadening stage $2^{-6}\lesssim Jt \lesssim 2^{-3}$.
In this window the initial-state return probability
$(1-p)^N$ decreases monotonically, so the recurrence
again occurs in min-relative NLM without a return of the quantum state.
The inset places this window within the full finite-system
dynamics, whose ordered spectrum and min-relative NLM have exact period
$\pi/(2J)$.
The sorted Gaussian limit and the controlled early-time
regime are established in Sec.~\ref{app:binomial-spectrum} of the Supplemental Material.

In the broadening regimes considered in Fig.~\ref{fig3},
corresponding revival times asymptotically double because
the spectral width grows linearly.
Their different amplitudes reflect their distinct
spectral shapes.
Together with the exponentially broadening squeezed state,
these examples illustrate the two roles:
the spectral shape $f$ fixes the min-relative NLM profile on the
logarithmic width axis, including its amplitude, while
the broadening law $L$ determines the recurrence times.

\textit{Conclusion and outlook.---}
We have shown that persistent min-relative NLM revivals can accompany continuous entanglement growth and a decreasing return probability under two-mode squeezing. For spectra approaching a fixed scaling shape, the fidelity branches at successive dyadic ranks become translated copies of a common profile on the logarithmic-width axis. Their optimized envelope therefore repeats as the spectral width doubles, with the spectral shape determining the asymptotic modulation and the broadening law governing its temporal recurrence. When the limiting profile is nonconstant, exponential and power-law broadening yield asymptotic periodicity in time and logarithmic time, respectively. Finite-dimensional squeezing reproduces the ideal revival pattern over a time window that grows with the local dimension. Beyond squeezing, number-conserving two-mode mixing verifies the predicted amplitude and asymptotic doubling of successive revival times in its early broadening regime at large particle number.

Our results establish a predictive link between entanglement-spectrum scaling and the temporal structure of min-relative NLM. A natural next question is how these revivals evolve when the rescaled spectral shape changes or develops multiple competing scales. Extending the analysis beyond a fixed limiting shape would clarify the stability and limits of the recurrence mechanism. Another direction is to identify interacting many-body dynamics that realize these spectral scaling regimes and test the predicted revival profiles both in numerical simulations and on quantum simulation platforms.

\textit{Acknowledgments.---}
GPT-6 Astra and GPT-6.2 Astra assisted with parts of manuscript proofreading and analytical derivations. Analytical results were verified independently by the authors. This work was supported by Key R\&D Program (Grant No. 2025YFE0217600), the National Natural Science Foundation of China (Grants Nos.U25A6009, T2121001, 92365301, 92565301).

\let\oldaddcontentsline\addcontentsline
\renewcommand{\addcontentsline}[3]{}%
\bibliography{main}
\let\addcontentsline\oldaddcontentsline
\onecolumngrid

\clearpage
\newpage

\widetext

\begin{center}
\textbf{\large Supplemental Material for \\``Persistent Nonlocal Magic Revivals from Entanglement-Spectrum Broadening''}
\end{center}

\addtocontents{toc}{\protect\setcounter{tocdepth}{0}}
{
\tableofcontents
}

\renewcommand{\theproposition}{S\arabic{proposition}}
\setcounter{proposition}{0}
\renewcommand{\thedefinition}{S\arabic{definition}}
\setcounter{definition}{0}

\renewcommand{\thefigure}{S\arabic{figure}}
\renewcommand{\theHfigure}{S\arabic{figure}}
\setcounter{figure}{0}
\renewcommand{\theequation}{S\arabic{equation}}
\renewcommand{\theHequation}{S\arabic{equation}}
\setcounter{equation}{0}
\renewcommand{\thesection}{\Roman{section}}

\setcounter{section}{0}
\renewcommand{\thetable}{S\arabic{table}}
\setcounter{table}{0}
\setcounter{secnumdepth}{4}

\section{Technical details of the two-mode squeezing model}
\label{append:squeezing}

\subsection{Entanglement entropy and spectral anti-flatness}

For the two-mode squeezed state Eq.~\ref{eq:squeezed-state}, the reduced state $\rho_A$ is diagonal
in the local Fock basis,
\begin{equation}
\rho_A=\sum_{m=0}^{\infty}\lambda_m|m\rangle\langle m|,
\end{equation}
where $\lambda_m=(1-q)q^m$ and $0\leq q<1$.
The geometric-series identities
\begin{align}
\sum_{m=0}^{\infty}q^m
&=\frac{1}{1-q},
\nonumber\\
\sum_{m=0}^{\infty}mq^m
&=q\frac{d}{dq}\frac{1}{1-q}
=\frac{q}{(1-q)^2}
\end{align}
give $\sum_m\lambda_m=1$ and the mean occupation
$\bar n=\sum_m m\lambda_m=q/(1-q)$.
The entanglement entropy therefore reads
\begin{align}
\mathcal S_A
&=-\sum_{m=0}^{\infty}\lambda_m\log_2\lambda_m
\nonumber\\
&=-\log_2(1-q)\sum_m\lambda_m
-\log_2 q\sum_m m\lambda_m
\nonumber\\
&=-\log_2(1-q)-\frac{q}{1-q}\log_2 q.
\label{eq:app-entropy}
\end{align}

The spectral moments likewise follow from a geometric sum:
\begin{align}
\operatorname{Tr}(\rho_A^\alpha)
&=(1-q)^\alpha\sum_{m=0}^{\infty}q^{\alpha m}
\nonumber\\
&=\frac{(1-q)^\alpha}{1-q^\alpha},
\qquad \alpha>0.
\label{eq:app-spectral-moments}
\end{align}
In particular,
\begin{align}
\operatorname{Tr}(\rho_A^2)&=\frac{1-q}{1+q},
\nonumber\\
\operatorname{Tr}(\rho_A^3)&=\frac{(1-q)^2}{1+q+q^2}.
\end{align}
Substituting these expressions into the definition of
anti-flatness gives
\begin{align}
\mathcal F_A
&=\operatorname{Tr}(\rho_A^3)
-[\operatorname{Tr}(\rho_A^2)]^2
\nonumber\\
&=(1-q)^2\left[
\frac{1}{1+q+q^2}-\frac{1}{(1+q)^2}
\right]
\nonumber\\
&=\frac{q(1-q)^2}{(1+q)^2(1+q+q^2)}.
\label{eq:app-antiflatness}
\end{align}

To obtain the large-$r$ asymptotics, take $r>0$, so that
$\bar n=\sinh^2r>0$. Using $q=\bar n/(1+\bar n)$,
we rewrite the entropy as
\begin{align}
\mathcal S_A
&=(\bar n+1)\log_2(\bar n+1)-\bar n\log_2\bar n
\nonumber\\
&=\log_2(\bar n+1)
+\bar n\log_2\left(1+\frac{1}{\bar n}\right).
\end{align}
As $r\to\infty$, we have $\bar n\sim e^{2r}/4$.
Using $\bar n+1=\cosh^2r$ for the first term and
$\ln(1+x)/x\to1$ as $x\to0$ for the second gives
\begin{align}
\log_2(\bar n+1)&=2r\log_2e-2+o(1),
\nonumber\\
\bar n\log_2\left(1+\frac{1}{\bar n}\right)
&=\log_2e+o(1).
\end{align}
Consequently,
\begin{equation}
\mathcal S_A=(2r+1)\log_2e-2+o(1),
\qquad r\to\infty.
\label{eq:app-entropy-asymptotics}
\end{equation}
For anti-flatness,
$1-q=\operatorname{sech}^2r\sim4e^{-2r}$,
while $q\to1$ and $(1+q)^2(1+q+q^2)\to12$.
Thus,
\begin{equation}
\mathcal F_A\sim\frac{(1-q)^2}{12}
\sim\frac43e^{-4r},
\qquad r\to\infty.
\label{eq:app-antiflatness-asymptotics}
\end{equation}

\subsection{Infinite-dimensional limit of min-relative nonlocal magic}
\label{app:infinite-limit}

For every fixed finite $r\ge0$, the bosonic state has infinitely many Schmidt coefficients,
whereas the stabilizer fidelity is initially defined on qubit
registers. We connect the two settings by a sequence of
normalized finite states. For $d=2^n$, write
\begin{equation}
|\Psi_n(r)\rangle
=\frac{1}{\sqrt{1-q^d}}
\sum_{m=0}^{d-1}\sqrt{(1-q)q^m}\,|m,m\rangle.
\label{eq:app-normalized-truncation}
\end{equation}
Each retained local Fock space is identified with an $n$-qubit
register. At fixed finite $r$,
\begin{equation}
|\langle\Psi(r)|\Psi_n(r)\rangle|^2
=1-q^d\longrightarrow1\qquad(n\to\infty).
\label{eq:app-truncation-overlap}
\end{equation}
State convergence alone does not establish convergence of an
optimization over growing stabilizer sets. For this step we
use the exact spectral characterization of locally optimized
stabilizer fidelity~\cite{sierant2026exactquantificationnonlocalmagic,viscardi2026nonlocalmagicclosedformsolution}.
Every bipartite pure qubit stabilizer state is locally
equivalent to Bell pairs and product spectators, so its
nonzero Schmidt probabilities are flat and have dyadic rank.
Consequently, the finite-register measure is
\begin{equation}
\mathcal M_A^{\mathrm{NL},(n)}(r)
=-\log_2
\frac{\displaystyle\max_{0\le k\le n}F_k(r)}
{1-q^{2^n}},
\label{eq:app-finite-nlm}
\end{equation}
where $F_k$ is defined using the untruncated probabilities in
Eq.~\ref{eq:fidelity-branches}.

To control the maximization, note that the full sum of square
roots is finite for $q<1$:
\begin{equation}
\left(\sum_{m=0}^{\infty}\sqrt{\lambda_m}\right)^2
=\frac{1+\sqrt q}{1-\sqrt q}=e^{2r}.
\end{equation}
It follows that $0\le F_k\le e^{2r}2^{-k}\to0$ as $k\to\infty$, whereas
$F_0=1-q>0$. The maximum is therefore attained at a finite
index. Once this index is included in the truncation,
Eq.~\ref{eq:app-finite-nlm} becomes
\begin{equation}
\mathcal M_A^{\mathrm{NL},(n)}(r)
=\mathcal M_A^{\mathrm{NL},\infty}(r)
+\log_2(1-q^{2^n}),
\end{equation}
which proves the fixed-$r$ limit
\begin{equation}
\mathcal M_A^{\mathrm{NL},\infty}(r)
:=-\log_2\max_{k\ge0}F_k(r).
\label{eq:app-infinite-nlm}
\end{equation}
At $r=0$, $F_k=2^{-k}$, and both the finite and infinite
measures vanish. The limit does not require a normalized
state at $q=1$: late-time behavior is studied only after
the infinite-register limit has been defined for each
finite $r$.

The construction is also consistent under changes of local
encoding and register size. Changes of encoding are absorbed
by the local-unitary optimization. Adding a local pure
spectator zero-pads the Schmidt spectrum. If its original
register dimension is $d$, all branches of rank at most $d$
remain unchanged, while branches of rank $R>d$ have the same
square-root sum as the rank-$d$ branch and a larger
denominator. They cannot improve the optimum. Thus adding
spectators does not change the finite measure, and the
sequence of embeddings defines a consistent spectral
extension. The reference resource here remains the qubit
stabilizer structure; the construction does not introduce
a separate continuous-variable stabilizer resource theory.

\subsection{Exact fidelity branches and optimal-sector switching}
\label{app:sector-switching}

For the geometric Schmidt probabilities, set $\tau=\tanh r=\sqrt q\in[0,1)$.
Geometric summation gives the exact branches
\begin{align}
F_k(r)
&=\frac{1-q}{2^k}
\left(\frac{1-\tau^{2^k}}{1-\tau}\right)^2
\nonumber\\
&=\frac{e^{2r}}{2^k}(1-\tau^{2^k})^2,
\label{eq:app-geometric-branches}
\end{align}
where $(1+\tau)/(1-\tau)=e^{2r}$. Consequently,
\begin{equation}
\frac{F_{k+1}}{F_k}
=\frac12\left(1+\tau^{2^k}\right)^2.
\label{eq:app-branch-ratio}
\end{equation}
For fixed $r>0$, this ratio strictly decreases with $k$ and
tends to $1/2$. Thus the branch sequence increases while the
ratio exceeds one and decreases after it falls below one.
Adjacent crossings therefore determine the global optimum.

The $k$ and $k+1$ sectors exchange optimality at
\begin{equation}
r_k=\operatorname{arctanh}
\left[(\sqrt2-1)^{2^{-k}}\right],\qquad k\ge0.
\label{eq:app-sector-crossings}
\end{equation}
These times are strictly increasing. Sector $0$ is optimal
for $0\le r\le r_0$, and sector $k\ge1$ is optimal for
$r_{k-1}\le r\le r_k$, with adjacent sectors tied at each
endpoint. At $r=0$, $F_k=2^{-k}$ and the optimum is sector
$0$. The upper envelope in Fig.~\ref{fig1}(d) therefore switches successively from one Bell-pair sector to the next.

To see the behavior at a crossing, write
$R_k=F_{k+1}/F_k$. For $m=2^k$ and $r>0$,
\begin{equation}
R_k'(r)=m\tau^{m-1}(1+\tau^m)(1-\tau^2)>0.
\end{equation}
At $r=r_k$, $R_k=1$, so differentiating
$F_{k+1}=F_kR_k$ gives
$F_{k+1}'-F_k'=F_kR_k'>0$.
The upper envelope is locally the maximum of the two
smooth branches, which agree in value but have different
slopes. It is therefore continuous but has a cusp.

\subsection{Asymptotic periodic profile and revival amplitude}
\label{app:periodicity}

For $\chi>0$ and $r>0$, let $\varepsilon=e^{-2r}\in(0,1)$ and $z=2^k\varepsilon$.
Since $\tau=(1-\varepsilon)/(1+\varepsilon)$,
Eq.~\ref{eq:app-geometric-branches} can be written exactly as
\begin{equation}
F_k(r)=g_\alpha(z),\qquad
g_\alpha(z)=\frac{(1-e^{-2\alpha z})^2}{z},\qquad
\alpha=\frac{\operatorname{arctanh}\varepsilon}{\varepsilon}.
\end{equation}
As $r\to\infty$, $\alpha=1+\varepsilon^2/3+O(\varepsilon^4)$.
The limiting branch is therefore
\begin{equation}
h(z)=g_1(z)=\frac{(1-e^{-2z})^2}{z}.
\label{eq:app-limiting-branch}
\end{equation}
To control the optimization, observe that
\begin{equation}
0\le\partial_\alpha g_\alpha(z)
=4e^{-2\alpha z}(1-e^{-2\alpha z})\le1.
\end{equation}
The mean-value theorem gives the uniform bound
\begin{equation}
0\le F_k(r)-h(2^k\varepsilon)
\le\alpha-1=O(e^{-4r}),
\label{eq:app-uniform-branch-error}
\end{equation}
valid for every $k\ge0$.

Writing $s=2r\log_2e$, define
\begin{equation}
A(s)=\max_{j\in\mathbb Z}h(2^{j-s}),\qquad
Q(s)=-\log_2A(s).
\label{eq:app-periodic-profile}
\end{equation}
The maximum exists because $h$ tends to zero at both ends
of the positive axis. For sufficiently large $r$, the
arguments added by extending $k\ge0$ to $j\in\mathbb Z$
are smaller than $\varepsilon$ and are dominated by the
$k=0$ branch, since $h$ increases near zero. Moreover, a
dyadic sample always lies in $[1,2]$, so
$A(s)\ge\min_{1\le z\le2}h(z)>0$ uniformly in $s$.
Equation~\ref{eq:app-uniform-branch-error} consequently
remains valid after maximization and taking the logarithm:
\begin{equation}
\mathcal M_A^{\mathrm{NL},\infty}(r)
=Q(2r\log_2e)+O(e^{-4r}).
\label{eq:app-periodic-asymptotics}
\end{equation}
Reindexing $j$ gives $Q(s+1)=Q(s)$. The uniform estimate
also controls the error at sector crossings.

The extrema follow from the shape of $h$. Its derivative is
\begin{equation}
h'(z)=\frac{1-e^{-2z}}{z^2}
\left[4ze^{-2z}-(1-e^{-2z})\right].
\end{equation}
It has a unique maximum at the positive solution
$z_m$ of $e^{2z_m}=1+4z_m$, with $z_m\simeq0.628$. The function $e^{2z}-1-4z$ is strictly convex, vanishes at zero, has negative initial slope, and diverges to infinity; hence it has exactly one additional zero on $(0,\infty)$.
The adjacent-sample ratio is
\begin{equation}
\frac{h(2z)}{h(z)}=\frac12(1+e^{-2z})^2,
\end{equation}
so the crossing occurs at
\begin{equation}
z_c=\frac12\ln(1+\sqrt2)\simeq0.441,
\qquad h(z_c)=h(2z_c).
\end{equation}
This ratio decreases with $z$. An optimal argument thus
satisfies $z_c\le z\le2z_c$; these inequalities express
dominance over both neighboring dyadic samples and,
by unimodality of the sample sequence, over all samples.
During one phase cycle, the winning argument moves from
$2z_c$ to $z_c$, passing through $z_m$. There is a unique
minimum phase modulo one, where a sample equals $z_m$.
Hence one is the fundamental period of $Q$, giving
\begin{equation}
\Delta r=\frac{\ln2}{2},\qquad
T=\frac{\ln2}{2\chi}.
\end{equation}
The limiting extrema and peak-to-trough amplitude are
\begin{align}
Q_{\min}&=-\log_2h(z_m)\simeq0.296,
\nonumber\\
Q_{\max}&=-\log_2h(z_c)\simeq0.361,
\nonumber\\
\Delta\mathcal M
&=\log_2\frac{h(z_m)}{h(z_c)}
\simeq0.065.
\label{eq:app-revival-amplitude}
\end{align}
These are the asymptotic bounds shown in Fig.~\ref{fig1}(c).
The NLM approaches this periodic profile; it is not exactly
periodic at finite $r$.

The recurrence does not imply a return of the state.
Indeed, $P_0(r)=\operatorname{sech}^2r$ obeys
$P_0'(r)=-2\operatorname{sech}^2r\tanh r<0$ for $r>0$.
Meanwhile, with $\bar n=\sinh^2r$,
\begin{equation}
\frac{d\mathcal S_A}{dr}
=\sinh(2r)\log_2\frac{\bar n+1}{\bar n}>0.
\end{equation}
The revived quantity is the optimized spectral matching
to the dyadic stabilizer reference family.

\subsection{Finite-dimensional dynamics and cutoff effects}
\label{app:finite-dynamics}

For Fig.~\ref{fig2}, each mode retains the Fock levels
$0,\ldots,d-1$, with $d=2^n$, and evolves under $H_d=P_dHP_d$.
Starting from the vacuum, conservation of the occupation
difference confines the dynamics to $|m,m\rangle$.
In this paired basis,
\begin{equation}
\frac{H_d}{\chi}
=i\sum_{m=0}^{d-2}(m+1)
\left(|m+1,m+1\rangle\langle m,m|
-|m,m\rangle\langle m+1,m+1|\right).
\end{equation}
Writing $|\Psi_d(r)\rangle=\sum_{m=0}^{d-1}c_m^{(d)}(r)|m,m\rangle$
gives
\begin{equation}
\frac{dc_m^{(d)}}{dr}
=m c_{m-1}^{(d)}-(m+1)c_{m+1}^{(d)},
\label{eq:app-finite-amplitudes}
\end{equation}
with $c_{-1}^{(d)}=c_d^{(d)}=0$ and
$c_m^{(d)}(0)=\delta_{m0}$. The upper boundary changes
the dynamics; this state differs from the normalized
projection used in Sec.~\ref{app:infinite-limit}.

After sorting the probabilities $|c_m^{(d)}|^2$ into
$\lambda_j^{(d)\downarrow}$, we evaluate
\begin{align}
\mathcal M_{A,d}^{\mathrm{NL}}(r)&=-\log_2\mathfrak F_d(r),
\nonumber\\
\mathfrak F_d(r)&=\max_{0\le k\le n}\frac1{2^k}
\left(\sum_{j=0}^{2^k-1}
\sqrt{\lambda_j^{(d)\downarrow}(r)}\right)^2.
\end{align}
Sorting is necessary because finite evolution need not
preserve the ordering of the Fock populations.
Numerically, the diagonal transformation $D_{mm}=i^m$
maps $H_d/\chi$ to a real symmetric tridiagonal matrix
$J_d=D^\dagger(H_d/\chi)D$, whose off-diagonal entries
are $(J_d)_{m,m+1}=m+1$ and whose diagonal vanishes.
If $J_d=V\operatorname{diag}(\omega_\alpha)V^{\mathsf T}$,
then
\begin{equation}
c_m^{(d)}(r)=i^m\sum_{\alpha=0}^{d-1}
V_{m\alpha}V_{0\alpha}e^{-i\omega_\alpha r}.
\end{equation}
This evaluates the finite matrix exponential directly.

To bound the cutoff error, let
$b_m(r)=\operatorname{sech}r\,\tanh^m r$ be the infinite
amplitudes and let $v(r)$ contain their first $d$ entries
without renormalization. For $K_d=-iH_d/\chi$, the only
boundary residual is
\begin{equation}
v'(r)-K_dv(r)=-d b_d(r)e_{d-1},
\end{equation}
where $e_{d-1}$ is the last coordinate vector.
Since $K_d$ is antisymmetric, its propagator preserves
the Euclidean norm. Duhamel's formula yields
\begin{align}
\|c^{(d)}(r)-v(r)\|_2&\le B_d(r),
\nonumber\\
B_d(r)&=d\int_0^r\operatorname{sech}u\,\tanh^d u\,du
\le\sinh r\,\tanh^d r.
\end{align}
The last inequality follows from
$d\operatorname{sech}u\,\tanh^d u
=\sinh u\,(d/du)\tanh^d u$ and monotonicity of $\sinh u$.
The omitted tail has squared norm $q^d$.
After zero-padding $c^{(d)}$ to the infinite space,
\begin{equation}
\|c^{(d)}(r)-b(r)\|_2
\le\sqrt{B_d(r)^2+q^d}
\le\epsilon_d(r):=\cosh r\,\tanh^d r.
\label{eq:app-finite-state-error}
\end{equation}

Taking absolute values and decreasingly sorting cannot
increase the Euclidean distance. Each dyadic root fidelity
is the inner product of the ordered amplitude magnitudes
with a unit vector that is constant on its first $2^k$
entries and zero elsewhere. Maximizing these inner products
is one-Lipschitz. Ranks above $d$ cannot improve the
zero-padded finite optimum, as shown in
Sec.~\ref{app:infinite-limit}. Therefore,
\begin{equation}
\left|\sqrt{\mathfrak F_d(r)}
-\sqrt{\mathfrak F_\infty(r)}\right|\le\epsilon_d(r),
\qquad \mathfrak F_\infty(r)=\max_{k\ge0}F_k(r).
\end{equation}
If $\epsilon_d<\sqrt{\mathfrak F_\infty}$, this implies
\begin{equation}
\left|\mathcal M_{A,d}^{\mathrm{NL}}(r)
-\mathcal M_A^{\mathrm{NL},\infty}(r)\right|
\le-2\log_2\left(1-
\frac{\epsilon_d(r)}{\sqrt{\mathfrak F_\infty(r)}}\right).
\label{eq:app-finite-nlm-error}
\end{equation}

For any fixed $R<\infty$, the state error is uniformly
bounded by $\cosh R\,\tanh^dR\to0$ on $0\le r\le R$.
The interval can also grow: set $R_d=\alpha\ln d$ with
$0<\alpha<1/2$. Using $\ln\tanh r\le-2e^{-2r}$ and
$\mathfrak F_\infty(r)\ge\operatorname{sech}^2r$ gives
\begin{equation}
\sup_{0\le r\le R_d}
\frac{\epsilon_d(r)}{\sqrt{\mathfrak F_\infty(r)}}
\le d^{2\alpha}\exp[-2d^{1-2\alpha}]
\longrightarrow0.
\end{equation}
Thus the NLM converges uniformly on an expanding time
window $0\le t\le\alpha\ln d/\chi$.
This establishes a controlled finite-dimensional
realization of successive squeezing revivals.
At fixed $d$, the bound does not control arbitrarily late
times, when the upper Fock boundary can modify the dynamics.

\section{Min-relative nonlocal magic revivals for broadening Schmidt spectra}
\subsection{Scaling form of broadening Schmidt spectra}
\label{app:broadening-spectra}

We consider a family of ordered Schmidt probability spectra,
\begin{equation}
\lambda_0^\downarrow(t)\ge\lambda_1^\downarrow(t)\ge\cdots\ge0,
\qquad
\sum_{j\ge0}\lambda_j^\downarrow(t)=1.
\label{eq:app-broadening-normalization}
\end{equation}
The parameter $t$ denotes the spectral evolution time and finite spectra are extended by zeros.

We focus on spectra that spread over an increasing number
of Schmidt coefficients while approaching a fixed profile
after rescaling the index and the probabilities.  We describe this shape by a measurable function $f$ satisfying:
\begin{enumerate}
\item[(i)] $f(u)\ge0$ for $u\ge0$;
\item[(ii)] $f$ is nonincreasing on $[0,\infty)$;
\item[(iii)] $\displaystyle\int_0^\infty f(u)\,du=1$.
\end{enumerate}
Let $L=L(t)>0$ be a spectral scale with $L(t)\to\infty$ as $t\to \infty$.
Informally, the scaling form is
\begin{equation}
\lambda_j^\downarrow(t)\simeq
\frac{1}{L(t)}f\!\left(\frac{j}{L(t)}\right).
\label{eq:app-spectral-scaling-heuristic}
\end{equation}
Thus the spectrum spreads over rank intervals of order $L$,
with a characteristic probability scale $L^{-1}$.
The scale $L$ need not be an integer or coincide with a moment
of the rank distribution.

To state the approximation precisely while preserving discrete
normalization, we define the cell-integrated spectrum
\begin{equation}
p_j^{(L)}
:=\int_{j/L}^{(j+1)/L}f(u)\,du,
\qquad j=0,1,\ldots .
\label{eq:app-cell-integrated-spectrum}
\end{equation}
The cells partition the positive axis, so
$\sum_{j\ge0}p_j^{(L)}=1$ exactly.
Monotonicity of $f$ also gives
$p_j^{(L)}\ge p_{j+1}^{(L)}$.
This construction does not require pointwise evaluation of $f$
at the origin and remains well defined for integrable
singularities.

We formulate this scaling behavior precisely by requiring
the evolving spectrum to approach $p^{(L(t))}$ in the
square-root distance:
\begin{equation}
\delta(t)^2
:=\sum_{j\ge0}
\left[
\sqrt{\lambda_j^\downarrow(t)}
-\sqrt{p_j^{(L(t))}}
\right]^2
\longrightarrow0
\qquad (t\to\infty).
\label{eq:app-spectral-scaling-distance}
\end{equation}
This condition controls the approximation over the entire
spectrum, including its tail. It provides a precise meaning
for the scaling form introduced above.

For two-mode squeezing, the Schmidt probabilities are
\begin{equation}
\lambda_j(t)=(1-q)q^j,
\qquad q=\tanh^2(\chi t).
\end{equation}
This spectrum is represented exactly by the exponential
profile and spectral scale
\begin{equation}
f(u)=e^{-u},
\qquad
L(t)=-\frac{1}{\ln q(t)}.
\label{eq:app-squeezing-profile-scale}
\end{equation}
Indeed,
\begin{equation}
p_j^{(L)}
=\int_{j/L}^{(j+1)/L}e^{-u}\,du
=(1-e^{-1/L})e^{-j/L}
=(1-q)q^j.
\end{equation}
Thus $\delta(t)=0$ for every $t>0$.
At late times,
\begin{equation}
L(t)\sim\frac14e^{2\chi t},
\end{equation}
so the spectral width grows exponentially while the
rescaled profile remains fixed.

\subsection{Limiting fidelity profile and scale periodicity}
\label{app:scale-profile}

Using the spectral expression for the min-relative NLM,
we write
\begin{align}
\mathcal M_A^{\mathrm{NL},\infty}(t)
&=-\log_2\max_{k\ge0}F_{2^k}(t),
\nonumber\\
F_R(t)
&=\frac1R
\left(\sum_{j=0}^{R-1}
\sqrt{\lambda_j^\downarrow(t)}\right)^2.
\label{eq:app-scale-fidelity-branches}
\end{align}
For a flat reference spectrum of rank $R$, the relative
rank $y=R/L(t)$ measures its support compared with the
width of the evolving spectrum. The corresponding
limiting fidelity profile is
\begin{equation}
H_f(y)
:=\frac1y
\left(\int_0^y\sqrt{f(u)}\,du\right)^2,
\qquad y>0.
\label{eq:app-limiting-fidelity-profile}
\end{equation}
Equivalently,
\begin{equation}
H_f(y)=\left|\langle\sqrt f,\phi_y\rangle\right|^2,
\qquad
\phi_y(u)=\frac{\mathbf1_{[0,y]}(u)}{\sqrt y}.
\label{eq:app-profile-overlap}
\end{equation}
Both $\sqrt f$ and $\phi_y$ have unit $L^2$ norm,
so $0\le H_f(y)\le1$.

To control the approximation uniformly over the candidate
ranks, introduce the step density associated with the
cell-integrated spectrum,
\begin{equation}
f_L(u)=Lp_j^{(L)},
\qquad
\frac jL\le u<\frac{j+1}{L}.
\label{eq:app-step-density}
\end{equation}
As $L\to\infty$, the cell width $1/L$ tends to zero.
Since $f$ is nonincreasing, the cell average $f_L(u)$
converges to $f(u)$ at every positive continuity point
of $f$, and hence almost everywhere. Both densities are normalized, so Scheff\'e's
lemma gives $\|f_L-f\|_1\to0$. Consequently,
\begin{equation}
\eta_L^2
:=\|\sqrt{f_L}-\sqrt f\|_2^2
\le\|f_L-f\|_1
\longrightarrow0.
\label{eq:app-profile-discretization-error}
\end{equation}

For every integer $R\ge1$, setting $y=R/L$ gives
the exact identity
\begin{equation}
\sqrt{F_R[p^{(L)}]}
=\frac1{\sqrt y}\int_0^y\sqrt{f_L(u)}\,du.
\label{eq:app-step-profile-identity}
\end{equation}
The normalized discrete flat vector, with entries
$R^{-1/2}$ for $j<R$ and zero otherwise, has unit norm.
Cauchy--Schwarz therefore bounds the error from replacing
the actual spectrum by $p^{(L(t))}$.
Applying the same argument to $\phi_y$ bounds the
remaining discretization error:
\begin{align}
\left|\sqrt{F_R(t)}
-\sqrt{F_R[p^{(L(t))}]}\right|
&\le\delta(t),
\nonumber\\
\left|\sqrt{F_R[p^{(L(t))}]}
-\sqrt{H_f(R/L(t))}\right|
&\le\eta_{L(t)}.
\label{eq:app-root-fidelity-error-components}
\end{align}
Thus
\begin{align}
\sup_{R\ge1}
\left|\sqrt{F_R(t)}
-\sqrt{H_f(R/L(t))}\right|
&\le\varepsilon(t),
\nonumber\\
\varepsilon(t)
:=\delta(t)+\eta_{L(t)}
&\longrightarrow0.
\label{eq:app-uniform-root-fidelity-error}
\end{align}
The bound is uniform in $R$, allowing the approximation
to remain valid after optimization over the reference rank.

We next control the endpoints of the continuous profile.
The function $H_f$ is continuous for $y>0$, and
\begin{equation}
H_f(y)\le\int_0^y f(u)\,du\longrightarrow0
\qquad (y\to0).
\label{eq:app-profile-small-rank}
\end{equation}
For $y>B>0$, splitting the integral at $B$ gives
\begin{equation}
\sqrt{H_f(y)}
\le
\frac1{\sqrt y}\int_0^B\sqrt{f(u)}\,du
+\left(\int_B^\infty f(u)\,du\right)^{1/2}.
\label{eq:app-profile-large-rank}
\end{equation}
Taking $y\to\infty$ and then $B\to\infty$ proves
$H_f(y)\to0$ as $y\to\infty$, without requiring
$\int_0^\infty\sqrt f(u)\,du$ to be finite.

Set $s=\log_2L$ and define
\begin{equation}
A_f(s):=\max_{\ell\in\mathbb Z}H_f(2^{\ell-s}),
\qquad
Q_f(s):=-\log_2A_f(s).
\label{eq:app-dyadic-limiting-envelope}
\end{equation}
The endpoint bounds ensure that the maximum is attained.
Moreover, every such dyadic grid contains a point in
$[1,2]$, giving the phase-independent lower bound
\begin{equation}
A_f(s)\ge b_f^2,
\qquad
b_f:=\min_{1\le y\le2}\sqrt{H_f(y)}>0.
\label{eq:app-envelope-positive-bound}
\end{equation}
The original optimization uses $k\ge0$.
The additional indices $\ell<0$ satisfy $2^\ell/L<1/L$,
and their fidelities obey
\begin{equation}
H_f(2^\ell/L)
\le\int_0^{1/L}f(u)\,du\longrightarrow0.
\label{eq:app-negative-sector-bound}
\end{equation}
They therefore cannot attain the maximum for sufficiently
large $L$.
Combining this observation with the uniform error bound
yields
\begin{equation}
\left|
\sqrt{\max_{k\ge0}F_{2^k}(t)}
-\sqrt{A_f(\log_2L(t))}
\right|
\le\varepsilon(t).
\label{eq:app-optimized-root-fidelity-error}
\end{equation}
For sufficiently large $t$, such that
$\varepsilon(t)<b_f$, taking logarithms gives
\begin{align}
&\left|
\mathcal M_A^{\mathrm{NL},\infty}(t)
-Q_f(\log_2L(t))
\right|
\nonumber\\
&\qquad\le
-2\log_2\left(1-\frac{\varepsilon(t)}{b_f}\right)
\longrightarrow0.
\label{eq:app-optimized-nlm-error}
\end{align}
Hence
\begin{equation}
\mathcal M_A^{\mathrm{NL},\infty}(t)
=Q_f(\log_2L(t))+o(1).
\label{eq:app-general-scale-asymptotics}
\end{equation}

Finally, reindexing the dyadic samples gives
\begin{equation}
A_f(s+1)
=\max_{\ell\in\mathbb Z}H_f(2^{\ell-s-1})
=A_f(s),
\end{equation}
and therefore
\begin{equation}
Q_f(s+1)=Q_f(s).
\label{eq:app-logwidth-periodicity}
\end{equation}
Doubling the spectral width and the reference rank leaves
their relative matching unchanged.
The limiting NLM profile consequently has a period of one
on the logarithmic-width axis.
This periodicity alone does not establish a nonzero
oscillation amplitude, since $Q_f$ may be constant.

\subsection{Conditions for nonconstant revivals and
quantitative predictions}
\label{app:revival-criterion}

The scale periodicity established above does not by itself
imply oscillations, since $Q_f$ may be constant.
We first characterize when the limiting profile is
nonconstant. Define
\begin{equation}
H_*:=\max_{y>0}H_f(y),
\qquad
K_f:=\{y>0:H_f(y)=H_*\}.
\label{eq:app-profile-maximum-set}
\end{equation}
The endpoint properties proved in subsection B ensure
that $K_f$ is nonempty.
At least one phase has a dyadic sample at a global
maximizer, so $\max_s A_f(s)=H_*$.
Consequently, $Q_f$ is constant if and only if every
phase samples at least one point of $K_f$:
\begin{equation}
K_f\cap\{2^{\ell-s}:\ell\in\mathbb Z\}\ne\varnothing
\qquad
\text{for every }s\in[0,1).
\label{eq:app-constant-profile-criterion}
\end{equation}
If this condition holds, then $A_f(s)=H_*$ at every phase.
If it fails, some phase has $A_f(s)<H_*$, whereas another
attains $H_*$, giving a nonconstant profile.

In particular, if $H_f$ has a unique global maximum
at $y_m$, the optimal value $H_*$ is attained only
at the phase
\begin{equation}
s_m=-\log_2y_m\pmod1.
\label{eq:app-optimal-matching-phase}
\end{equation}
Every other phase misses $y_m$, so $Q_f$ is nonconstant.
The same conclusion holds whenever $K_f$ is finite.
Conversely, a global-maximum plateau spanning $[a,2a]$
is sampled at every phase and therefore produces
a constant profile.

Such a plateau is possible even for a continuous,
strictly decreasing spectral shape.
For example, consider
\begin{equation}
f(u)=C
\begin{cases}
(3/2-u)^2,&0\le u\le1,\\
1/(4u),&1<u\le2,\\
e^{-(u-2)}/8,&u>2,
\end{cases}
\qquad
C=\left(\frac{29}{24}+\frac{\ln2}{4}\right)^{-1}.
\label{eq:app-constant-profile-example}
\end{equation}
Direct integration gives
\begin{equation}
H_f(y)=C
\begin{cases}
y(3/2-y/2)^2,&0<y\le1,\\
1,&1<y\le2,\\
\bigl(3-e^{-(y-2)/2}\bigr)^2/(2y),&y>2.
\end{cases}
\label{eq:app-constant-profile-example-fidelity}
\end{equation}
The first branch increases to $C$, and the last decreases
from $C$.
Thus $[1,2]$ is a global-maximum plateau, and
$Q_f(s)=-\log_2C$ is constant.
Spectral broadening with a strictly decreasing shape
is therefore insufficient, by itself, to guarantee
nonzero oscillations.

We next obtain explicit predictions when $H_f$ is
strictly unimodal: it is strictly increasing on $(0,y_m)$
and strictly decreasing on $(y_m,\infty)$.
On the interval $y_m/2<y<y_m$, the function
\begin{equation}
D(y):=H_f(y)-H_f(2y)
\end{equation}
is strictly increasing, with
\begin{equation}
D(y_m/2)<0,\qquad D(y_m)>0.
\end{equation}
There is therefore a unique
$y_c\in(y_m/2,y_m)$ satisfying
\begin{equation}
H_f(y_c)=H_f(2y_c).
\label{eq:app-adjacent-profile-crossing}
\end{equation}

At any phase, only the two neighboring dyadic samples
surrounding $y_m$ can maximize the fidelity:
all smaller samples lie farther down the increasing
branch, and all larger samples lie farther down
the decreasing branch.
The dominant reference rank changes when these two
candidates cross at $y_c$.
For the sector of rank $2^k$, the interval of optimality is
\begin{equation}
\frac{2^{k-1}}{y_c}\le L\le\frac{2^k}{y_c}.
\label{eq:app-profile-sector-window}
\end{equation}
As $L$ increases across this interval, its relative rank
$2^k/L$ decreases from $2y_c$ to $y_c$, passing through
$y_m$.
The optimized fidelity first increases to $H_f(y_m)$
and then decreases to $H_f(y_c)$.
Hence
\begin{equation}
Q_{f,\min}=-\log_2H_f(y_m),
\qquad
Q_{f,\max}=-\log_2H_f(y_c),
\label{eq:app-profile-extrema}
\end{equation}
and the peak-to-trough amplitude is
\begin{equation}
\Delta\mathcal M_f
=\log_2\frac{H_f(y_m)}{H_f(y_c)}>0.
\label{eq:app-general-revival-amplitude}
\end{equation}
There is exactly one minimum phase modulo one,
so the fundamental period of $Q_f$ is one.

The corresponding minima and maxima of the limiting
profile occur at spectral widths
\begin{equation}
L_k^{\min}=\frac{2^k}{y_m},
\qquad
L_k^{\max}=\frac{2^k}{y_c}.
\label{eq:app-profile-extrema-widths}
\end{equation}
These locations describe the limiting profile;
finite-width corrections can shift the extrema
of the actual NLM.

To convert these scale predictions into times, suppose
that $L(t)$ is continuous and eventually strictly increasing.
Define the predicted times by $L(t_k^a)=L_k^a$,
where $a\in\{\min,\max\}$, $y_{\min}=y_m$,
and $y_{\max}=y_c$.
For exponential broadening,
\begin{equation}
L(t)\sim L_0e^{\gamma t},
\qquad \gamma>0,
\end{equation}
the predicted times satisfy
\begin{equation}
t_k^a
=\frac{k\ln2-\ln(L_0y_a)}{\gamma}+o(1),
\label{eq:app-exponential-extrema-times}
\end{equation}
so
\begin{equation}
t_{k+1}^a-t_k^a
\longrightarrow\frac{\ln2}{\gamma}.
\label{eq:app-exponential-time-period}
\end{equation}
For power-law broadening,
\begin{equation}
L(t)\sim L_0(t/t_0)^\nu,
\qquad \nu>0,
\end{equation}
they instead satisfy
\begin{equation}
    t_k^a=t_0 \left(\frac{2^k}{L_0y_a}\right)^{1/\nu}[1+o(1)],
\label{eq:app-powerlaw-extrema-times}
\end{equation}
\begin{equation}
\frac{t_{k+1}^a}{t_k^a} \longrightarrow2^{1/\nu}.
\label{eq:app-powerlaw-extrema-times2}
\end{equation}
Thus the spectral shape determines the limiting amplitude
and extrema phases, whereas the broadening law determines
their temporal spacing.
For two-mode squeezing, subsection A gives $\gamma=2\chi$,
recovering the asymptotic time period $\ln2/(2\chi)$.

\subsection{Inverse-cubic spectra and logarithmic-time revivals}
\label{app:inverse-cubic-spectrum}

We consider the normalized Schmidt spectrum
\begin{equation}
\lambda_j(L)=\frac{(j+L)^{-3}}{\zeta(3,L)},
\qquad
\zeta(3,L)=\sum_{j=0}^{\infty}(j+L)^{-3},
\label{eq:app-inverse-cubic-spectrum}
\end{equation}
with $L=vt>0$ and $v>0$.
This example prescribes the spectrum directly.
Its limiting shape is
\begin{equation}
f_{\mathrm P}(u)=\frac{2}{(1+u)^3}.
\label{eq:app-inverse-cubic-shape}
\end{equation}

To verify the scaling condition in subsection A, define
\begin{equation}
a_j^{(L)}=\frac1L f_{\mathrm P}(j/L),
\qquad
S_L=\sum_{j\ge0}a_j^{(L)}=2L^2\zeta(3,L).
\end{equation}
Then $\lambda_j(L)=a_j^{(L)}/S_L$.
Since $f_{\mathrm P}$ is nonincreasing and normalized,
comparison of the left-endpoint sum with its integral gives
\begin{equation}
1\le S_L\le1+\frac2L.
\label{eq:app-inverse-cubic-normalization-bound}
\end{equation}
Moreover, the cell-integrated spectrum
\begin{equation}
p_{j,\mathrm P}^{(L)}
=\int_{j/L}^{(j+1)/L}f_{\mathrm P}(u)\,du
\end{equation}
satisfies $a_j^{(L)}\ge p_{j,\mathrm P}^{(L)}$.
Hence
\begin{align}
&\sum_j\left|\lambda_j(L)-p_{j,\mathrm P}^{(L)}\right|
\nonumber\\
&\quad\le
\sum_j|\lambda_j(L)-a_j^{(L)}|
+\sum_j|a_j^{(L)}-p_{j,\mathrm P}^{(L)}|
\nonumber\\
&\quad=2(S_L-1)\le\frac4L.
\label{eq:app-inverse-cubic-lone-error}
\end{align}
The square-root distance therefore obeys
\begin{equation}
\delta_{\mathrm P}(L)^2
\le\frac4L\longrightarrow0.
\label{eq:app-inverse-cubic-scaling-error}
\end{equation}
Thus this spectrum satisfies the assumptions of
subsections A and B.

The limiting fidelity profile is
\begin{equation}
H_{\mathrm P}(y)
=\frac8y
\left(1-\frac1{\sqrt{1+y}}\right)^2.
\label{eq:app-inverse-cubic-fidelity-profile}
\end{equation}
Writing $x=\sqrt{1+y}>1$, this becomes
\begin{equation}
H_{\mathrm P}(y)=\frac{8(x-1)}{x^2(x+1)}.
\end{equation}
Its derivative with respect to $x$ has the sign of
$-x^2+x+1$.
Consequently, $H_{\mathrm P}$ is strictly unimodal, with
\begin{equation}
y_{m,\mathrm P}=\frac{1+\sqrt5}{2}.
\label{eq:app-inverse-cubic-maximizer}
\end{equation}
The unique crossing point determined by
\begin{equation}
H_{\mathrm P}(y_{c,\mathrm P})
=H_{\mathrm P}(2y_{c,\mathrm P})
\end{equation}
is $y_{c,\mathrm P}\simeq1.146$.
Subsection C then gives
\begin{align}
Q_{\mathrm P,\min}&\simeq0.471,
\nonumber\\
Q_{\mathrm P,\max}&\simeq0.508,
\nonumber\\
\Delta\mathcal M_{\mathrm P}
&\simeq0.037.
\label{eq:app-inverse-cubic-extrema}
\end{align}

For $L=vt$, the NLM therefore satisfies
\begin{equation}
\mathcal M_{\mathrm P}(t)
=Q_{\mathrm P}(\log_2vt)+o(1),
\qquad t\to\infty.
\label{eq:app-inverse-cubic-nlm-asymptotics}
\end{equation}
The predicted minima and maxima of the limiting profile
occur at
\begin{equation}
\widehat t_k^{\min}
=\frac{2^k}{v y_{m,\mathrm P}},
\qquad
\widehat t_k^{\max}
=\frac{2^k}{v y_{c,\mathrm P}}.
\label{eq:app-inverse-cubic-predicted-times}
\end{equation}
Corresponding predicted times double between successive
cycles.
The limiting profile is therefore periodic in $\log_2t$,
with a fundamental period of one.

\subsection{Number-conserving two-mode mixing}
\label{app:binomial-spectrum}

We now consider the number-conserving Hamiltonian
\begin{equation}
H=J(a^\dagger b+ab^\dagger),
\qquad J>0,
\label{eq:app-mixing-hamiltonian}
\end{equation}
with initial state $|N,0\rangle$ and integer $N\ge1$.
We take $N+1=2^n$, so each local occupation space can
be encoded in $n$ qubits.
The exact evolution is
\begin{equation}
|\Psi_N(t)\rangle
=\sum_{m=0}^N
\sqrt{\binom Nm}
[\cos(Jt)]^{N-m}[-i\sin(Jt)]^m
|N-m,m\rangle.
\label{eq:app-mixing-exact-state}
\end{equation}
Its Schmidt probabilities are
\begin{equation}
\lambda_m^{(N)}(t)
=\binom Nm p(t)^m[1-p(t)]^{N-m},
\qquad
p(t)=\sin^2(Jt).
\label{eq:app-mixing-binomial-probabilities}
\end{equation}
Here $m$ labels the occupation of mode $B$;
the descending Schmidt order is denoted below by $j$.

Let
\begin{equation}
\mu=Np,\qquad
\sigma^2=Np(1-p).
\end{equation}
We first establish the scaling limit along sequences
for which $\sigma\to\infty$, allowing $p$ to depend on $N$.
Define the discrete Gaussian approximation
\begin{equation}
g_m=\frac1{\sqrt{2\pi}\sigma}
\exp\!\left[-\frac{(m-\mu)^2}{2\sigma^2}\right],
\qquad 0\le m\le N.
\label{eq:app-binomial-discrete-gaussian}
\end{equation}
For every fixed $M>0$, Stirling's formula gives
\begin{equation}
\lambda_m^{(N)}=g_m[1+o(1)]
\end{equation}
uniformly on $|m-\mu|\le M\sigma$.
Both boundaries lie asymptotically far from the mean
in units of $\sigma$.
Combining the central approximation with the binomial
variance bound on the tails, and then taking
$M\to\infty$, yields
\begin{equation}
\sum_{m=0}^N|\lambda_m^{(N)}-g_m|
\longrightarrow0.
\label{eq:app-binomial-gaussian-lone}
\end{equation}

Decreasing rearrangement cannot increase the $\ell^1$
distance, so
\begin{equation}
\sum_{j=0}^N
|\lambda_j^{(N)\downarrow}-g_j^\downarrow|
\longrightarrow0.
\label{eq:app-sorted-binomial-gaussian-lone}
\end{equation}
The Gaussian values are ordered by increasing distance
from $\mu$.
On central rank ranges, the distance associated with
the $j$-th ordered value satisfies
\begin{equation}
d_j=\frac j2+O(1),
\end{equation}
because an interval extending a distance $d$ on each
side of the mean contains $2d+O(1)$ lattice points.
Consequently, for $j=O(\sigma)$,
\begin{equation}
g_j^\downarrow
=\frac1{\sqrt{2\pi}\sigma}
\exp\!\left[-\frac{j^2}{8\sigma^2}\right][1+o(1)].
\label{eq:app-sorted-gaussian-profile}
\end{equation}
This identifies
\begin{equation}
L(N,t)=\sqrt8\,\sigma
=\sqrt{8Np(1-p)},
\qquad
f_{\mathrm G}(u)=\frac2{\sqrt\pi}e^{-u^2}.
\label{eq:app-binomial-profile-scale}
\end{equation}
The compact-rank approximation and Gaussian tail control
imply
\begin{equation}
\sum_{j\ge0}
\left|
g_j^\downarrow-
\int_{j/L}^{(j+1)/L}f_{\mathrm G}(u)\,du
\right|
\longrightarrow0,
\label{eq:app-sorted-gaussian-cell-convergence}
\end{equation}
where finite vectors are extended by zeros.
Together with the rearrangement bound, this proves
the square-root convergence condition of subsection A
for the sorted binomial spectrum.

The limiting fidelity profile is
\begin{equation}
H_{\mathrm G}(y)
=\frac{\sqrt\pi}{y}
\operatorname{erf}^2\!\left(\frac y{\sqrt2}\right),
\label{eq:app-gaussian-fidelity-profile}
\end{equation}
where
\begin{equation}
\operatorname{erf}(z)
=\frac2{\sqrt\pi}\int_0^z e^{-u^2}\,du.
\end{equation}
To verify strict unimodality, write
\begin{align}
I_{\mathrm G}(y)
&=\int_0^y\sqrt{f_{\mathrm G}(u)}\,du,
\nonumber\\
B_{\mathrm G}(y)
&=2y\sqrt{f_{\mathrm G}(y)}-I_{\mathrm G}(y).
\end{align}
The derivative of $H_{\mathrm G}$ has the sign of
$B_{\mathrm G}$, while
\begin{equation}
B_{\mathrm G}'(y)
=\sqrt{f_{\mathrm G}(y)}(1-2y^2).
\end{equation}
Thus $B_{\mathrm G}$ starts at zero, first increases
and then strictly decreases to a negative limit.
It has one positive zero, establishing strict unimodality.
The maximum and crossing positions are
\begin{equation}
y_{m,\mathrm G}\simeq1.400
\qquad
y_{c,\mathrm G}\simeq0.976.
\label{eq:app-gaussian-profile-critical-scales}
\end{equation}
The resulting limiting extrema and amplitude are
\begin{align}
Q_{\mathrm G,\min}&\simeq0.168,
\nonumber\\
Q_{\mathrm G,\max}&\simeq0.290,
\nonumber\\
\Delta\mathcal M_{\mathrm G}
&\simeq0.122.
\label{eq:app-gaussian-profile-extrema}
\end{align}

The linear-broadening regime is the joint limit
\begin{equation}
N\to\infty,\qquad
Jt\to0,\qquad
\sqrt N\,Jt\to\infty,
\label{eq:app-mixing-joint-limit}
\end{equation}
or equivalently $N^{-1/2}\ll Jt\ll1$.
In this window,
\begin{equation}
L(N,t)\sim\sqrt{8N}\,Jt,
\end{equation}
and
\begin{equation}
\mathcal M_{A,N}^{\mathrm{NL}}(t)
=
Q_{\mathrm G}\!\left(
\log_2[\sqrt{8N}\,Jt]
\right)+o(1).
\label{eq:app-mixing-logtime-nlm}
\end{equation}
The large-width limit is therefore taken jointly in
$N$ and $t$, rather than at arbitrarily late times
for fixed $N$.

On the initial broadening branch $0\le Jt\le\pi/4$,
the exact width is
\begin{equation}
L(N,t)=\sqrt{2N}\sin(2Jt).
\end{equation}
The limiting-profile predictions for the extrema
consequently give
\begin{equation}
J\widehat t_{k,N}^{a}
=\frac12\arcsin\!\left(
\frac{2^k}{\sqrt{2N}\,y_a}
\right),
\label{eq:app-mixing-predicted-times}
\end{equation}
whenever the argument lies in $[0,1]$,
with $y_{\min}=y_{m,\mathrm G}$ and
$y_{\max}=y_{c,\mathrm G}$.
In the joint regime
\begin{equation}
1\ll2^k\ll\sqrt N,
\label{eq:app-mixing-cycle-window}
\end{equation}
this reduces to
\begin{equation}
J\widehat t_{k,N}^{a}
\sim\frac{2^k}{\sqrt{8N}\,y_a}.
\label{eq:app-mixing-early-predicted-times}
\end{equation}
Corresponding predicted times therefore have ratio
tending to two, provided both remain within this window.

Finally, the initial-state return probability is
\begin{equation}
P_0(t)=[1-p(t)]^N,
\end{equation}
which decreases monotonically during the early
broadening stage.
Separately, $p(t+\pi/(2J))=1-p(t)$ only reverses the
occupation labels of the binomial spectrum.
The ordered spectrum and NLM therefore repeat under
$t\mapsto t+\pi/(2J)$.
This exact finite-system periodicity is distinct from
the logarithmic-time modulation during initial
spectral broadening.

\section{Nonlocal stabilizer R\'enyi entropy under two-mode squeezing}
\label{app:nlsre-squeezing}
\subsection{Stabilizer R\'enyi entropy, nonlocal SRE,
and the reference-state conjecture}
\label{app:nlsre-definition}

For a pure $N$-qubit state $|\Psi\rangle$, the stabilizer
R\'enyi entropy (SRE)~\cite{PhysRevLett.128.050402} is defined as
\begin{equation}
\mathcal M_2(|\Psi\rangle)
=-\log_2\zeta_2(|\Psi\rangle),\qquad
\zeta_2(|\Psi\rangle)
=\frac{1}{2^N}\sum_{P\in\mathcal P_N}
|\langle\Psi|P|\Psi\rangle|^4,
\label{eq:app-sre2-definition}
\end{equation}
where $\mathcal P_N=\{I,X,Y,Z\}^{\otimes N}$ is the set of
$4^N$ Pauli strings.
The SRE vanishes on pure stabilizer states, is invariant
under Clifford unitaries, and is additive on tensor products.

Consider a bipartition $A{:}B$ with $n_A$ and $n_B$ qubits,
respectively, so that $N=n_A+n_B$ and $d_X=2^{n_X}$ for
$X=A,B$. The nonlocal SRE (NL-SRE)~\cite{Qian_2025} is
\begin{equation}
\mathcal M_{A,2}^{\mathrm{NL}}(|\Psi\rangle)
:=\min_{U_A\in U(d_A),\,U_B\in U(d_B)}
\mathcal M_2\!\left[(U_A\otimes U_B)|\Psi\rangle\right].
\label{eq:app-nlsre-definition}
\end{equation}
The optimization runs over arbitrary unitaries on the
complete local subsystems.

The Schmidt decomposition across the $A{:}B$ cut can be written as
\begin{equation}
|\Psi\rangle
=\sum_{m=0}^{d-1}\sqrt{\lambda_m}
|a_m\rangle_A|b_m\rangle_B,
\qquad d=\min(d_A,d_B),
\label{eq:app-sre-schmidt-decomposition}
\end{equation}
where the probabilities are ordered as
$\lambda_0\ge\lambda_1\ge\cdots\ge\lambda_{d-1}\ge0$,
$\sum_m\lambda_m=1$, and zero Schmidt values are included
when the Schmidt rank is smaller than $d$.
The reference state associated with the ordered Schmidt
values is defined on a
$\log_2 d=\min(n_A,n_B)$ qubits system as
\begin{equation}
|\psi_{\mathrm{ref}}(\boldsymbol\lambda)\rangle
:=\sum_{m=0}^{d-1}\sqrt{\lambda_m}\,|m\rangle,
\qquad d=\min(d_A,d_B),
\end{equation}
where $|m\rangle$ is the computational-basis vector
whose binary label represents the integer $m$.
The reference-state optimality conjecture asserts that
\begin{equation}
\mathcal M_{A,2}^{\mathrm{NL}}(|\Psi\rangle)
=\mathcal M_2\!\left(
|\psi_{\mathrm{ref}}(\boldsymbol\lambda)\rangle\right).
\end{equation}
We define
\begin{equation}
\mathcal M_{A,2}^{\mathrm{ref}}(\boldsymbol\lambda)
:=\mathcal M_2\!\left(
|\psi_{\mathrm{ref}}(\boldsymbol\lambda)\rangle\right).
\label{eq:app-reference-sre-definition}
\end{equation}
Local changes of Schmidt basis bring \(|\Psi\rangle\) to the bipartite representative \(|C_{\boldsymbol\lambda}\rangle=\sum_m\sqrt{\lambda_m}|m\rangle_A|m\rangle_B\). A CNOT from each active qubit of \(A\) to its partner in \(B\) maps this representative to \(|\psi_{\rm ref}(\boldsymbol\lambda)\rangle\) on the active qubits of \(A\), with all other qubits in \(|0\rangle\). Clifford invariance and additivity of the stabilizer Rényi entropy therefore give \(\mathcal M_2(|C_{\boldsymbol\lambda}\rangle)=\mathcal M_2(|\psi_{\rm ref}(\boldsymbol{\lambda})\rangle)\). Since \(|C_{\boldsymbol\lambda}\rangle\) belongs to the local-unitary orbit of \(|\Psi\rangle\), its SRE provides an upper bound on the minimum over that orbit. Hence,
\begin{equation}
0\le\mathcal M_{A,2}^{\mathrm{NL}}(|\Psi\rangle)
\le\mathcal M_{A,2}^{\mathrm{ref}}(\boldsymbol\lambda).
\label{eq:app-sre-reference-bound}
\end{equation}

\textit{Reference-state optimality conjecture.---}
The minimum in Eq.~\eqref{eq:app-nlsre-definition} is
conjectured to be attained by the ordered reference state~\cite{liu2026nonlocalnonstabilizernessslightlyentangled,torre2026nonlocalmagicentanglementspectrum}:
\begin{equation}
\mathcal M_{A,2}^{\mathrm{NL}}(|\Psi\rangle)
=\mathcal M_{A,2}^{\mathrm{ref}}(\boldsymbol\lambda)
=\mathcal M_2\!\left(
|\psi_{\mathrm{ref}}(\boldsymbol\lambda)\rangle\right).
\label{eq:app-sre-reference-conjecture}
\end{equation}
We distinguish this conjectured equality from the
unconditional upper bound in
Eq.~\eqref{eq:app-sre-reference-bound}.
In the following subsections, we evaluate the reference
SRE for the normalized geometric Schmidt spectrum and
establish its asymptotic revivals. Under the
reference-state optimality conjecture, these results
also describe the locally optimized NL-SRE.

\subsection{Exact reference-state SRE for the truncated geometric spectrum}

For the two-mode squeezed state, retain $d=2^n$ Schmidt
levels and normalize their probabilities as
\begin{equation}
\lambda_m^{(n)}(r)
=\frac{(1-q)q^m}{1-q^{2^n}},\qquad
0\le m<2^n,\qquad q=\tanh^2r.
\label{eq:app-sre-geometric-spectrum}
\end{equation}
The corresponding $n$-qubit reference state and its SRE are
\begin{equation}
|\psi_{n,\mathrm{ref}}(r)\rangle
=\sqrt{\frac{1-q}{1-q^{2^n}}}
\sum_{m=0}^{2^n-1}q^{m/2}|m\rangle,
\qquad
\mathcal M_{A,2}^{\mathrm{ref},(n)}(r)
:=\mathcal M_2(|\psi_{n,\mathrm{ref}}(r)\rangle).
\label{eq:app-sre-geometric-reference}
\end{equation}
At $r=0$, the state is the vacuum and its reference SRE is zero. For $r>0$, introduce $x=-\ln q>0$ and write the binary
label as $m=\sum_{j=0}^{n-1}2^jb_j$, with $b_j\in\{0,1\}$.
The identity
\begin{equation}
\prod_{j=0}^{n-1}(1+q^{2^j})
=\sum_{m=0}^{2^n-1}q^m
=\frac{1-q^{2^n}}{1-q}
\label{eq:app-sre-binary-normalization}
\end{equation}
follows by expanding the product: each choice of binary
digits contributes exactly one power $q^m$.
It gives the factorized probabilities
\begin{equation}
\lambda_m^{(n)}
=\prod_{j=0}^{n-1}
\frac{e^{-2^jxb_j}}{1+e^{-2^jx}}.
\label{eq:app-sre-binary-probabilities}
\end{equation}
Taking square roots therefore yields the exact state
factorization
\begin{equation}
|\psi_{n,\mathrm{ref}}(x)\rangle
=\bigotimes_{j=0}^{n-1}|\omega(2^jx)\rangle,
\qquad
|\omega(y)\rangle
:=\frac{|0\rangle+e^{-y/2}|1\rangle}
{\sqrt{1+e^{-y}}}.
\label{eq:app-sre-binary-product}
\end{equation}
Here $|\psi_{n,\mathrm{ref}}(x)\rangle$ denotes the same
reference state parametrized by $x$.
This factorization is compatible with the prescribed
descending order of the geometric spectrum.

For the single-qubit factor $|\omega(y)\rangle$, the
Pauli expectation values are
\begin{equation}
\langle X\rangle_y=\operatorname{sech}(y/2),\qquad
\langle Y\rangle_y=0,\qquad
\langle Z\rangle_y=\tanh(y/2).
\label{eq:app-sre-single-qubit-bloch}
\end{equation}
Its stabilizer purity is consequently
\begin{align}
\zeta_2(|\omega(y)\rangle)
&=\frac12\left[
1+\operatorname{sech}^4(y/2)+\tanh^4(y/2)
\right]\nonumber\\
&=1-\tanh^2(y/2)+\tanh^4(y/2),
\label{eq:app-sre-single-qubit-purity}
\end{align}
where $\operatorname{sech}^2(y/2)+\tanh^2(y/2)=1$ was used.
The single-qubit SRE contribution is defined by
\begin{equation}
g_2(y):=-\log_2\!\left[
1-\tanh^2(y/2)+\tanh^4(y/2)
\right].
\label{eq:app-sre-single-qubit-function}
\end{equation}
By the additivity of SRE on tensor products,
Eq.~\eqref{eq:app-sre-binary-product} gives
\begin{equation}
\zeta_2(|\psi_{n,\mathrm{ref}}(x)\rangle)
=\prod_{j=0}^{n-1}\zeta_2(|\omega(2^jx)\rangle).
\end{equation}
Taking its negative logarithm proves the exact
finite-register formula
\begin{equation}
\mathcal{M}_{A,2}^{\mathrm{ref},(n)}(r)
=\sum_{j=0}^{n-1}g_2\!\left(2^jx(r)\right),
\qquad x(r)=-\ln\tanh^2r.
\label{eq:app-sre-finite-sum}
\end{equation}
Thus the reference SRE is a sum of contributions sampled
at successive dyadic scales. 

\subsection{Infinite-dimensional limit and asymptotic periodicity}
\label{app:sre-asymptotic-periodicity}

\begin{figure}[t]
\centering
 \includegraphics[width=0.8\linewidth]{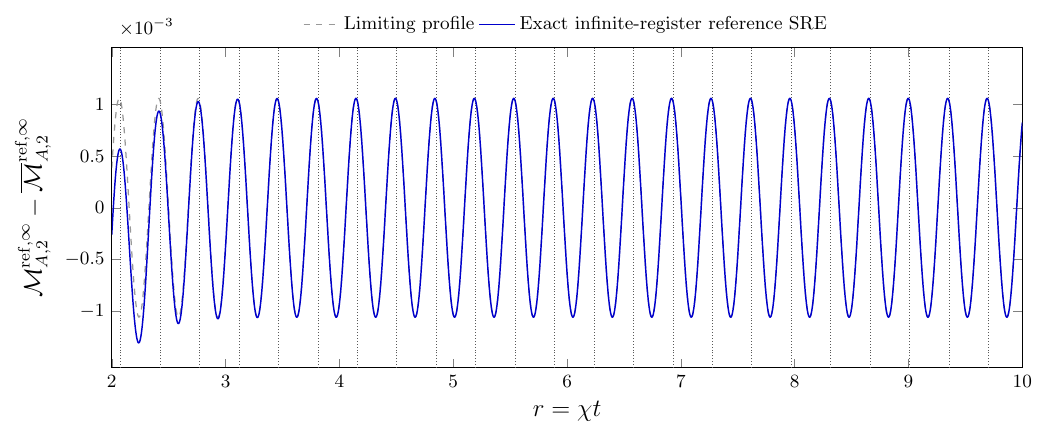}
\caption{Asymptotic revivals of the reference SRE under
two-mode squeezing. The blue curve shows
$\mathcal M_{A,2}^{\mathrm{ref},\infty}(r)
-\overline{\mathcal M}_{A,2}^{\mathrm{ref},\infty}$
as a function of $r=\chi t$, where
$\overline{\mathcal M}_{A,2}^{\mathrm{ref},\infty}$
is the mean of the asymptotic periodic profile.
The oscillations approach a nonzero amplitude.
The vertical dashed lines are separated by
$\Delta r=\ln2/2$, corresponding to the asymptotic
revival period $T=\ln2/(2\chi)$.}
\label{appfig:refnlm}
\end{figure}

We first take $n\to\infty$ at fixed $r>0$, or equivalently
at fixed $x=-\ln\tanh^2r>0$.
To control the dyadic sum in
Eq.~\eqref{eq:app-sre-finite-sum}, we define
\begin{equation}
u(y)=\tanh^2(y/2)\operatorname{sech}^2(y/2),
\qquad g_2(y)=-\frac{\ln[1-u(y)]}{\ln2}.
\end{equation}
For $y>0$, we have $0\le u(y)\le1/4$,
$u(y)\le y^2/4$, and $u(y)\le4e^{-y}$.
Using $-\ln(1-u)\le4u/3$ therefore gives
\begin{equation}
0\le g_2(y)
\le\frac{\min\{y^2,16e^{-y}\}}{3\ln2}.
\label{eq:app-sre-kernel-bound}
\end{equation}
The exponential bound ensures convergence at every fixed
$x>0$, so the infinite-dimensional reference SRE is
well defined by
\begin{equation}
\mathcal M_{A,2}^{\mathrm{ref},\infty}(r)
:=\lim_{n\to\infty}
\mathcal M_{A,2}^{\mathrm{ref},(n)}(r)
=F_2(x(r)),\qquad
F_2(x):=\sum_{j=0}^{\infty}g_2(2^jx).
\label{eq:app-sre-infinite-sum}
\end{equation}
More explicitly, with $d=2^n$, the omitted terms obey
\begin{align}
0\le F_2(x)-\mathcal M_{A,2}^{\mathrm{ref},(n)}(r)
&\le\frac{16}{3\ln2}
\sum_{\ell=0}^{\infty}e^{-2^\ell dx}
\nonumber\\
&\le\frac{16e^{-dx}}{3\ln2(1-e^{-dx})},
\label{eq:app-sre-truncation-tail}
\end{align}
where $x=x(r)$ and the final inequality follows from
$2^\ell\ge\ell+1$.

To expose the asymptotic scale dependence, extend the sum
to all integer dyadic scales and define
\begin{equation}
P_2(s):=\sum_{j\in\mathbb Z}g_2(2^{j-s}).
\label{eq:app-sre-periodic-profile}
\end{equation}
For $s\in[0,1]$, the terms with $j<0$ are bounded by a
constant times $4^j$, while those with $j\ge0$ are bounded
by a constant times $e^{-2^{j-1}}$.
Thus Eq.~\eqref{eq:app-sre-kernel-bound} gives uniform
convergence over one period. Reindexing the absolutely
convergent series yields
\begin{equation}
P_2(s+1)=P_2(s).
\label{eq:app-sre-scale-periodicity}
\end{equation}
At $s=-\log_2x$, the difference between this periodic
profile and the original one-sided sum is
\begin{equation}
P_2(-\log_2x)-F_2(x)
=\sum_{j=-\infty}^{-1}g_2(2^jx).
\end{equation}
The small-argument bound in
Eq.~\eqref{eq:app-sre-kernel-bound} gives
\begin{equation}
0\le P_2(-\log_2x)-F_2(x)
\le\frac{x^2}{3\ln2}\sum_{j=-\infty}^{-1}4^j
=\frac{x^2}{9\ln2}.
\label{eq:app-sre-profile-error}
\end{equation}
Consequently,
\begin{equation}
F_2(x)=P_2(-\log_2x)+O(x^2),\qquad x\to0^+.
\label{eq:app-sre-log-scale-asymptote}
\end{equation}
Combining the two tail bounds also gives a finite-dimensional
estimate,
\begin{equation}
0\le P_2(-\log_2x)
-\mathcal M_{A,2}^{\mathrm{ref},(n)}(r)
\le\frac{x^2}{9\ln2}
+\frac{16e^{-dx}}{3\ln2(1-e^{-dx})}.
\label{eq:app-sre-finite-profile-error}
\end{equation}
The periodic profile is therefore approached when $x$ is
small while the retained dimension satisfies $dx\gg1$.

For two-mode squeezing, $r=\chi t$ and
\begin{equation}
x(r)=-\ln\tanh^2r
=4\operatorname{arctanh}(e^{-2r})
=4e^{-2r}\left[1+O(e^{-4r})\right].
\label{eq:app-sre-squeezing-scale}
\end{equation}
It follows that
$-\log_2x(r)=2r\log_2e-2+O(e^{-4r})$.
To control the resulting phase error, note that
$g_2'(y)=O(y)$ as $y\to0$ and
$g_2'(y)=O(e^{-y})$ as $y\to\infty$.
The differentiated series
$P_2'(s)=-(\ln2)\sum_{j\in\mathbb Z}
2^{j-s}g_2'(2^{j-s})$
therefore also converges uniformly for $s\in[0,1]$,
and $P_2'$ is bounded.
Using this bound, Eq.~\eqref{eq:app-sre-profile-error},
and the periodicity of $P_2$ to remove the integer shift
$-2$, we obtain
\begin{equation}
\mathcal M_{A,2}^{\mathrm{ref},\infty}(r)
=P_2(2r\log_2e)+O(e^{-4r}).
\label{eq:app-sre-time-asymptote}
\end{equation}
A unit shift of the profile argument corresponds to
\begin{equation}
\Delta r=\frac{\ln2}{2},\qquad
T=\frac{\Delta r}{\chi}=\frac{\ln2}{2\chi}.
\label{eq:app-sre-asymptotic-period}
\end{equation}
This establishes a periodic limiting profile with a
controlled late-time error.

\end{document}